\documentclass{aastex61}
\pdfoutput=1 
\usepackage{amsmath,amstext}
\usepackage[T1]{fontenc}
\usepackage{apjfonts} 
\usepackage[figure,figure*]{hypcap}

\usepackage{color}  
\usepackage[figuresright]{rotating}

\newcommand{\glc}{Galacticus}
\newcommand{\cloudy}{{\sc Cloudy}}

\newcommand{\wfirst}{{\it WFIRST}}

\newcommand{\romanM}{{\it Roman}}
\newcommand{\euclid}{{\it Euclid}}

\def\fun#1#2{\lower3.6pt\vbox{\baselineskip0pt\lineskip.9pt
        \ialign{$\mathsurround=0pt#1\hfill##\hfil$\crcr#2\crcr\sim\crcr}}}

\shorttitle{Roman HLSS}
\shortauthors{Wang et al.}

\begin{document}

\title{The High Latitude Spectroscopic Survey on the Nancy Grace Roman Space Telescope}

\author{Yun Wang}\footnote{*wang@ipac.caltech.edu}
\affiliation{IPAC, California Institute of Technology, Mail Code 314-6, 1200 E. California Blvd., Pasadena, CA 91125}
\author{Zhongxu Zhai}
\affiliation{IPAC, California Institute of Technology, Mail Code 314-6, 1200 E. California Blvd., Pasadena, CA 91125}
\author{Anahita Alavi}
\affiliation{IPAC, California Institute of Technology, Mail Code 314-6, 1200 E. California Blvd., Pasadena, CA 91125}
\author{Elena Massara}
\affiliation{Waterloo Centre for Astrophysics, University of Waterloo, 200 University Ave W, Waterloo ON N2L 3G1, Canada}
\affiliation{Department of Physics and Astronomy, University of Waterloo, Waterloo, ON N2L 3G1, Canada}
\author{Alice Pisani}
\affiliation{Princeton University, Department of Astrophysical Sciences, 4 Ivy Lane, Princeton, 08540, New Jersey, USA}
\affiliation{Center for Computational Astrophysics, Flatiron Institute, 162 5th Avenue, New York, NY, 10010, USA}
\affiliation{The Cooper Union for the Advancement of Science and Art, 41 Cooper Square, New York, NY 10003, USA}
\author{Andrew Benson}
\affiliation{Carnegie Observatories, 813 Santa Barbara Street, Pasadena, CA 91101}
\author{Christopher M. Hirata}
\affiliation{Department of Physics, The Ohio State University, Columbus, OH 43210}
\affiliation{Department of Astronomy, The Ohio State University, Columbus, OH 43210}
\affiliation{Center for Cosmology and AstroParticle Physics, The Ohio State University, Columbus, OH 43210}
\author{Lado Samushia}
\affiliation{Department of Physics, Kansas State University, 116 Cardwell Hall, 1228 N. 17th street, Manhattan, KS 66506, USA}
\author{David H. Weinberg}
\affiliation{Department of Astronomy, The Ohio State University, Columbus, OH 43210}
\affiliation{Center for Cosmology and AstroParticle Physics, The Ohio State University, Columbus, OH 43210}
\author{James Colbert}
\affiliation{IPAC, California Institute of Technology, Mail Code 314-6, 1200 E. California Blvd., Pasadena, CA 91125}
\author{Olivier Dor\'e} 
\affiliation{Jet Propulsion Laboratory, California Institute of Technology, Pasadena, CA 91101, USA}
\affiliation{Cahill Center for Astronomy and Astrophysics, California Institute of Technology, Pasadena, CA 91125, USA}
\author{Tim Eifler}
\affiliation{Department of Astronomy/Steward Observatory, University of Arizona, 933 North Cherry Avenue, Tucson, AZ 85721-0065, USA}
\author{Chen Heinrich}
\affiliation{Cahill Center for Astronomy and Astrophysics, California Institute of Technology, Pasadena, CA 91125, USA}
\author{Shirley Ho}
\affiliation{Center for Computational Astrophysics, Flatiron Institute, 162 5th Avenue, New York, NY, 10010, USA}
\author{Elisabeth Krause}
\affiliation{Department of Astronomy/Steward Observatory, University of Arizona, 933 North Cherry Avenue, Tucson, AZ 85721-0065, USA}
\author{Nikhil Padmanabhan}
\affiliation{Department of Physics, Yale University, New Haven, CT 06520, USA}
\author{David Spergel}
\affiliation{Center for Computational Astrophysics, Flatiron Institute, 162 5th Avenue, New York, NY, 10010, USA}
\author{Harry I. Teplitz}
\affiliation{IPAC, California Institute of Technology, Mail Code 314-6, 1200 E. California Blvd., Pasadena, CA 91125}

\begin{abstract}

The {\it Nancy Grace Roman Space Telescope} will conduct a High Latitude Spectroscopic Survey (HLSS) over a large volume at high redshift, using the near-IR grism (1.0-1.93 $\mu$m, $R=435-865$) and the 0.28 deg$^2$ wide field camera. We present a reference HLSS which maps 2000 deg$^2$ and achieves an emission line flux limit of 10$^{-16}$ erg/s/cm$^2$ at 6.5$\sigma$, requiring $\sim$0.6 yrs of observing time. We summarize the flowdown of the \romanM\ science objectives to the science and technical requirements of the HLSS. We construct a mock redshift survey over the full HLSS volume by applying a semi-analytic galaxy formation model to a cosmological N-body simulation, and use this mock survey to create pixel-level simulations of 4 deg$^2$ of HLSS grism spectroscopy. We find that the reference HLSS would measure $\sim$ 10 million H$\alpha$ galaxy redshifts that densely map large scale structure at $z=1-2$ and 2 million [OIII] galaxy redshifts that sparsely map structures at $z=2-3$. We forecast the performance of this survey for measurements of the cosmic expansion history with baryon acoustic oscillations and the growth of large scale structure with redshift space distortions. We also study possible deviations from the reference design, and find that a deep HLSS at $f_{\rm line}>7\times10^{-17}$erg/s/cm$^2$ over 4000 deg$^2$ (requiring $\sim$1.5 yrs of observing time) provides the most compelling stand-alone constraints on dark energy from \romanM\ alone. This provides a useful reference for future optimizations. The reference survey, simulated data sets, and forecasts presented here will inform community decisions on the final scope and design of the \romanM\ HLSS.
\end{abstract}


\section{Introduction}
\label{sec:intro}

The 2010 decadal survey of astronomy and astrophysics \citep{Astro2010}
identified a {\it Wide Field Infrared Survey Telescope} (\wfirst) as the
highest priority for a large new space initiative, recognizing that 
advances in large format near-IR detectors enabled a mission that could
make major contributions to studies of cosmic acceleration, 
exoplanet demographics, and a wide range of stellar and galactic
astrophysics reaching from the solar neighborhood to redshifts $z>10$.
A key component of the proposed \wfirst\ observing program was a 
large area slitless spectroscopic survey at high Galactic latitude,
designed to map large scale structure traced by galaxies over an 
enormous comoving volume at $z>1$.  This paper describes a ``reference''
version of this High Latitude Spectroscopic Survey (HLSS) based on
the current capabilities of the mission.  A companion paper
\citep{Hirata2021} describes a reference version of the
High Latitude Imaging Survey (HLIS).
The HLSS as described here has been defined as a concrete example of the
spectroscopic component of the High Latitude Wide Area Survey, in order to
formulate mission science requirements and specify needed mission capabilities.
The final implementation of the High Latitude Wide Area Spectroscopic Survey
will be defined by NASA through a community process prior to launch.

Like other decadal missions before it, \wfirst\ has undergone substantial
changes between the concept proposed to the decadal survey and the
design adopted for construction.  A crucial early change was the shift
from a purpose-built 1.5-m telescope to a pre-existing 2.4-m telescope,
with correspondingly higher light-gathering power and angular resolution
\citep{Dressler2012}.  A second crucial change was the adoption of
H4RG detectors, with $4\times$ the pixel count of the H2RG detectors
assumed in the original design.  A third change, prompted by the
greater capabilities of the 2.4-m telescope, was the addition of an 
on-axis coronagraphic instrument to study nearby exoplanetary systems
and demonstrate technology for future missions.  The evolution of 
the \wfirst\ concept and design can be traced through the series
of Science Definition Team reports 
\citep{Green2012,Spergel2013,Spergel2015}.

\wfirst\ is now in its final design and fabrication phase
(Phase C in NASA parlance), and it has been renamed the
{\it Nancy Grace Roman Space Telescope} in honor of one of the
pioneers of NASA space astrophysics.  We will refer to it hereafter
as {\it Roman Space Telescope}, or simply as \romanM.  
The mission recently passed a critical milestone of identifying a
full complement of flight-qualified H4RG detectors that meet its
science requirements.  While some detailed design decisions are still
to be made, the capabilities of the mission hardware are now 
well understood. 
In particular, the \romanM\ grism enables spectroscopy at $R=460\lambda/\mu$m for $\lambda=1-1.93\mu$m. 
The \romanM\ capabilities are summarized succinctly in the white paper
of \cite{Akeson2019}.  As emphasized there, for applications in 
wide field near-IR imaging and spectroscopy \romanM\ is many hundreds
of times more powerful than {\it Hubble Space Telescope}, enabling
ambitious observing programs that were previously inconceivable.

During its 5-year prime mission, \romanM\ will include a substantial
Guest Observer program, but it is expected that the majority of 
observing time will be devoted to a set of large, coherent surveys
as envisioned by Astro2010.
The design of these surveys will be guided by quantitative goals
in cosmology and exoplanet science, but the powerful public data sets
they produce will support an enormous range of science investigations.
Allocations of observing time and choices among alternative survey
strategies will be made close to mission launch by a community-driven
process.  Here we present a reference design for the HLSS, one that
illustrates the many considerations that go into survey strategy choices
and that demonstrates the extraordinary power of an observing program
that would (in this reference design) require $\sim 0.6$ years
of observing time (including calibration and overheads) on \romanM, interleaved throughout the prime mission.

In round numbers, this reference survey would map 2000 deg$^2$ to a 
typical emission-line flux limit of $10^{-16}$erg\,s$^{-1}$cm$^{-2}$ at $6.5\sigma$ (8.5$\times 10^{-17}$erg\,s$^{-1}$cm$^{-2}$ at $5\sigma$),\footnote{We base our "6.5$\sigma$" and "5$\sigma$" characterizations on the number of photons in the signal compared to the total amount of noise (which includes sky background, read noise, etc).  Because the noise depends on the total amount of exposure time (and how it is divided into multiple exposures) at a given point on the sky, the ratio of limiting fluxes is not simply 6.5/5.}
enabling measurement of $\sim$10M galaxy redshifts at $z=1-2$ using H$\alpha$ emission lines and $\sim$2M galaxy redshifts at $z=2-3$ using [OIII] emission lines.
We assume that all area observed by the HLSS will also be observed 
by the HLIS, providing high-resolution angular images and broad-band
spectral energy distributions (SEDs) for all detected galaxies.
For studies of cosmic acceleration, this galaxy redshift survey would
enable precise measurements of the distance-redshift relation $D_A(z)$ and the expansion rate $H(z)$ using the standard ruler imprinted by baryon acoustic oscillations (BAO, \citealt{Blake2003,Seo2003}),
and precise measurements of the growth of dark matter clustering through redshift-space distortions (RSD; \citealt{Kaiser1987,Guzzo2008,Wang_2008}).
General discussion of these observational probes in the context of
cosmic acceleration can be found in the relevant reviews, e.g.,
\cite{Frieman2008}, \cite{Wang2010book}, and \cite{Weinberg2013}.  Our forecasts in
this paper draw heavily on the methodology developed by
\cite{Wang2010,Wang2012,Wang2013,Wang2017} and \cite{Zhai_2019}.
As an illustration for broader applications in extragalactic astrophysics, we note that \romanM\ is capable of carrying out a deep galaxy survey over 20 deg$^2$ in $\sim$ 5 days, with the same depth as the influential near-IR grism survey of the 3D-HST Treasury Program (5$\sigma$ line flux limits of 5$\times 10^{-17}$ erg/s/cm$^2$ for "typical objects", \citealt{Brammer2012,VanDokkum2013,Momcheva2016}),
but over an the area more than $100$ times larger!

The HLSS envisioned here has complementary properties to the large galaxy
redshift surveys anticipated from the ground-based Dark Energy Spectroscopic
Instrument (DESI; \citealt{DESI2016}) and the space-based \euclid\ mission
\citep{Euclid}, both of which aim to cover larger areas (14,000 deg$^2$ and
15,000 deg$^2$, respectively).  Compared to DESI, where the primary DESI target class transitions from luminous red galaxies to emission-line galaxies at $z>1$, the HLSS will achieve a sampling of large scale structure $>10\times$ denser at $z =1.3$.
Compared to \euclid, the HLSS is smaller in
volume but much denser in sampling, with an estimated difference of a
factor of >5 in galaxy density at $z=1.5$.  

Although \romanM\ could execute a shallow and wide area survey comparable
to \euclid's in approximately one year of observing time, the deeper
survey proposed here is a better complement to other surveys and more
effectively exploits the capabilities of \romanM's larger aperture.
The higher density of galaxies with H$\alpha$ redshifts at $z=1-2$ allows
much better measurements of higher order clustering beyond 2-point correlations,
and the greater depth allows [OIII] redshift measurements
of $\sim$2M galaxies at $z=2-3$, a unique probe of structure in this redshift
range.  Per unit observing time, \romanM\ is an extraordinarily efficient
facility for slitless spectroscopic surveys, so it is well positioned
to respond to developments in experimental cosmology between now and
mission launch in the mid-2020s.

\section{Survey Definition}

\subsection{Science goal, objectives, and requirements}
\label{subsec:sciencegoals}

The Science Goal of the \romanM\ High Latitude Survey (HLS) is to answer the two top-level questions on dark energy:\\
(1) Is cosmic acceleration caused by a new energy component or by the breakdown of General Relativity (GR) on cosmological scales?\\
(2) If the cause is a new energy component, is its energy density constant in space and time, or has it evolved over the history of the Universe?\\

This Science Goal flows down to the following \romanM\ Science Objectives:\\
\noindent
Objective 1: \romanM\ will conduct near-infrared (NIR) sky surveys in both imaging and spectroscopic modes, providing an imaging sensitivity for unresolved sources better than 26.5 AB magnitude.\\
\noindent
Objective 2: \romanM\ will determine the expansion history of the Universe in order to test possible explanations of its apparent accelerating expansion, including dark energy and modification to Einstein’s gravity, using the supernova, weak lensing, and galaxy redshift survey techniques, at redshifts up to $z = 2$ with high-precision cross-checks between the techniques.\\
\noindent
Objective 3: \romanM\ will determine the growth history of the largest structures in the Universe in order to test the possible explanations of its apparent accelerating expansion including dark energy and modification to Einstein’s gravity using weak lensing, redshift space distortions, and galaxy cluster techniques, at redshifts up to $z =2$ with high-precision cross-checks between the techniques.\\

We have flowed down these three Science Objectives into Level 2 Science Requirements for the \romanM\ High Latitude Spectroscopic Survey (HLSS), as described below. 

{\bf \paragraph{HLSS 1} The area to be surveyed shall be $\sim$1500 deg$^2$ (2000 deg$^2$ goal) after correcting for edge effects.  This area will be contiguous to the extent practical, and at least 90\% of the survey area must also be covered by the high latitude imaging survey.}

 The survey area of $\sim$1500 deg$^2$ is the minimum required to yield robust BAO/RSD measurements. The $>90\%$ overlap with the HLIS enables joint analysis of 90\% of WL and GRS data, which could maximize the dark energy science from \romanM. Imaging also provides undispersed galaxy positions, required for robust redshift determination. The statistical precision of the dark energy
constraints is sensitive to the survey area as well as the survey depth of the HLSS; a trade study of depth versus area will need to be carried out to optimize both, in the context of combination with the weak lensing data from the \romanM\ HLIS, and synergy with Euclid and LSST. \cite{Eifler21b} provides an example of such a study. A survey of one or two large, contiguous areas has smaller edge effects and better window functions than a survey comprised of many smaller areas. A single contiguous area is preferable for the standard BAO analysis \citep{Dore18}.  

{\bf \paragraph{HLSS 2} The comoving density of galaxies with measured redshifts shall satisfy $n > 3\times10^{-4}\ (h/\textrm{Mpc})^3$ at $z=1.6$.}

This is set by requiring that the condition for shot-noise non-dominance, $nP_{0.2} \sim1$ (where $P_{0.2}$ is the galaxy power spectrum at wavenumber $k=0.2\,h/$Mpc), is met at $z=1.6$, with 20\% margin. Requiring $nP_{0.2}
\sim1$ implies $n> 3\times10^{-4}\ (h/\mathrm{Mpc})^3$ at $z=1.3$, and $n >6.5\times 10^{-4} (h/\mathrm{Mpc})^3$ at $z=1.8$. Given the forecast of H$\alpha$ emission line galaxy (ELG) counts by \cite{Zhai_2019} for a H$\alpha$ line flux limit of $f_{\rm line} > 10^{-16} \, \mathrm{ erg\,s^{-1}cm^{-2}}$, 
$nP_{0.2}\sim0.6$ at $z=1.8$, and $nP_{0.2}>2$ at $z=1.3$. 
We have assumed a bias for H$\alpha$ ELGs of $b(z) = 1+0.5z$. This is 
more conservative than the predictions from \cite{Zhai_2021} for a H$\alpha$ line flux limit of $f_{\rm line} > 10^{-16} \, \mathrm{ erg\,s^{-1}cm^{-2}}$.
We cannot require a higher galaxy number density than what nature provides, given fixed observing time and area coverage. Here we have chosen a characteristic high redshift, $z=1.6$, the end of the range where DESI can look for ELGs via the [O II] doublet \citep{DESI2019}.
We have allowed significant margin to allow for the large uncertainties in the measured H$\alpha$ LF due to the limited availability of uniform data. 
The \cite{Zhai_2019} H$\alpha$ LF we have assumed is consistent with Model 3 in \cite{Pozzetti:2016}.
The H$\alpha$ line flux limit of $f_{\rm line} > 10^{-16} \, \mathrm{ erg\,s^{-1}cm^{-2}}$, at which this science requirement is met, 
is significantly deeper than the \euclid\ Galaxy Redshift Survey (GRS), which ensures that the HLSS is deep enough for carrying out robust modeling of systematic effects for BAO/RSD, higher order statistics, and the combination of weak lensing and RSD as tests of GR. 

{\bf \paragraph{HLSS 3} The wavelength range of the HLSS will allow measurement of
H$\alpha$ emission line redshifts over the redshift range $1.1<z<1.9$.}

The redshift range $1.1<z<1.9$ corresponding wavelength range is 1.38 $\mu$m to 1.9 $\mu$m.  This wavelength coverage also allows measurements of [OIII] emission line redshifts over the range $1.8 < z < 2.8$.  
However, redshift measurement strongly depends on the spectroscopic design. For grism spectroscopy, at least two emission lines are required for a robust redshift determination. Choosing the wavelength range to be 1-1.93$\mu$m enables the detection of both H$\alpha$ and [OIII] over $1.1<z<1.9$, and both [OIII] and [OII] over $1.8 < z < 2.8$, with sufficient margins.
The key consideration is that a space mission should focus on what cannot be
accomplished from the ground, and be complementary to other space missions in
wavelength coverage. Ground-based galaxy redshift surveys can reach $z\sim1$ without great difficulty, thus we should focus on $z>1$. 
It is also critical that the \romanM\ HLSS redshift range is complementary to that of \euclid, with its red cutoff at 1.85 $\mu$m, or $z < 1.8$ for H$\alpha$. 
\euclid\ GRS can only reach $z \sim2$; its shallow depth does not enable a high enough number density of observed [OIII] ELGs. The HLSS is deep enough to observe both H$\alpha$ (656.3nm) and [OIII] (500nm) ELGs, with the number density of the latter sensitive to the survey depth (the deeper the survey the higher their number density).

{\bf \paragraph{HLSS 4} Redshift measurement errors $\sigma_z$ shall satisfy $\sigma_z < 0.001(1+z)$, excluding galaxies larger than $0.54''$ in radius and outliers. The fraction of outliers with $|z_\mathrm{obs}- z_\mathrm{true}|/(1+z_\mathrm{true})>0.005$ shall be less than 10\% in the sample of galaxies smaller than $0.54''$ in radius. The incidence of outliers shall be known to a fraction of $2\times 10^{-3}$ of the full sample at each redshift.}

The rms redshift error of $\sigma_z < 0.001(1+z)$ is sufficient to prevent degradation of dark energy measurements \citep{Wang2010}.
Outliers in the redshift measurement are contaminants to both BAO and RSD measurements, as they reduce the S/N of the BAO measurement, and bias the RSD measurement.
Adding a contaminant to the galaxy power spectrum with contamination
fraction $\alpha$ corresponds to leaking in an amount of power from the ``wrong'' line with amplitude $\alpha^2$, and diluting the power spectrum of the ``real''
signal by a factor of $(1-\alpha)^2$ at the same time. 
For BAO the dilution is a minor issue (it reduces S/N), but for RSD it is a critical problem, because it reduces the galaxy bias $b$ by
$(1-\alpha)$ without changing the linear redshift-space distortion parameter
$\beta$. So the inferred rate of growth of structure $f_g = b \beta $ is
reduced by a factor of $1-\alpha$.  This leads to a stringent requirement on
knowledge of $\alpha$. If the true $\alpha$ is 9.8\%, but estimated to be 10\%,
then the systematic error is 0.2\%, which is a reasonable budget for this contribution.

Larger size galaxies have larger redshift errors in grism spectroscopy. 
The current data from the HST grism survey WISP finds that 90\% of galaxies that would be observed by \romanM\ (H$\alpha$ line flux $> 10^{-16} \mathrm{erg/s/cm}^2$, $1 < z < 1.94$) have a size less than $0.54^{\prime\prime}$ (semi-major axis continuum size) \citep{WISP}. Since the maximum redshift of the WISP sample is $\sim1.6$, \romanM\ H$\alpha$ ELGs will likely have smaller sizes on average.
The [OIII] ELGs are more compact, so this requirement is also sufficient for
the redshift precision for [OIII] ELGs.

{\bf \paragraph{HLSS 5} Relative position measurement uncertainties shall be less than $3.4^{\prime\prime}$ over the entire survey area.}

In order to measure galaxy clustering accurately, galaxy positions are required to be measured to better than $\sim0.1 \mathrm{Mpc}/h$ (which corresponds to $6.9^{\prime\prime}$, assuming that 105 $\mathrm{Mpc/}h$ subtends $\sim$ 2 degrees, e.g., the Planck 2018 cosmology at $z=1.5$). 
Given the pixel scale of $0.11^{\prime\prime}$, this requirement is automatically met within each field, and is tied to the precision of astrometry across different fields.

{\bf \paragraph{HLSS 6} The survey completeness shall be 60\%, and the redshift purity shall be 90\% (i.e., the outlier fraction is less than 10\%), excluding galaxies larger than $0.54^{\prime\prime}$ in radius. Completeness is defined as the fraction of H$\alpha$ ELGs with measured redshifts flagged as reliable, and purity is defined as the fraction of measured redshifts flagged as reliable that are actually within 2$\sigma$ of the true redshifts.}

A requirement on completeness and purity is needed to translate the H$\alpha$ ELG number counts predicted by the H$\alpha$ LF to the galaxy number density that can be used to measure BAO/RSD by \romanM\ HLSS. The completeness of 60\% and redshift purity of 90\% are based on extrapolations from \euclid, and are being validated using grism simulations.  Since \romanM\ has a higher spatial and spectral resolution compared to \euclid, and more rolls (4 versus 3) per field, we expect a higher completeness and purity for \romanM. 
The requirement on the knowledge on the contamination fraction is set by HLSS 4.\\

These six science requirements lead to a number of implementation and operations requirements, see \cite{Dore18}. We will omit those here, except for the key requirements that impact the HLSS survey design.

{\bf \paragraph{HLSS Imaging} Imaging observations shall be obtained of the fields in the HLSS that reach JAB=24.0, HAB=23.5, and F184AB=23.1 for an $r_\mathrm{eff}=0.3^{\prime\prime}$ source at 10$\sigma$ to achieve a reference image position, in 3 filters.}

This requirement is met automatically for the area covered by both the HLSS and the HLIS, thus only needs to be applied to any area not covered by the HLIS.  Imaging in at least three filters is required to build a minimal spectral template for grism spectral decontamination.

{\bf \paragraph{HLSS Deep Field} There shall be 40 observations of two deep fields, each 11 deg$^2$ in area, sufficient to characterize the completeness and purity of the overall galaxy redshift sample. The 40 observations repeat the HLSS observing sequence of 4 exposures 10 times, with each deep field observation having the same exposure time as a wide field observation of the HLSS. The dispersion directions of the 40 observations should be roughly evenly distributed between 0 and 360 degrees.}

This requirement is derived from the need to calibrate the HLSS with a spectroscopic subsample with nearly 100\% purity. For meaningful comparisons, this subsample should have the same selection criteria as that of the HLSS, but with a redshift purity $>99\%$. To measure the redshift purity to 1\% requires 10,000 objects, assuming noise of $1/\sqrt{N}$ from Poisson statistics. This needs to be done in at least four categories (low $z$, high $z$, faint, luminous), hence we need 160,000 galaxies, to allow for up to 16 subdivisions.
Based on the estimated galaxy number density of $\sim 7273$ per deg$^2$ at the flux limit for the HLSS, $10^{-16} \mathrm{erg} \, \mathrm{s}^{-1}\mathrm{cm}^{-2}$, we need a total area for the deep fields of 160,000/7273=22 deg$^2$.  These can be split into two subfields of 11 deg$^2$ each.  Smaller subfields prevent the testing of galaxy clustering statistics in each subfield. Each deep field should be part of the HLSS footprint, so they are representative of the HLSS as a whole.

The visits to the deep field should consist of 10 sets of HLSS-like visits,
matching the integration time, dither pattern, and observational time-sequence
of the HLSS strategy, with each set of HLSS-like visits covering the same
areas of 22 deg$^2$. Assuming an efficiency (completeness multiplied by purity) of 50\% and uncorrelated sets, the efficiency after 10 sets of visits is (1-0.5)$^{10}=0.001$, leading to a 99.9\% complete sample for calibrating the HLSS. Since each set of observation consists of 4 roll angles, the total number of deep field observations is 40. The dispersion directions of the 40 visits should be roughly evenly distributed between 0 and 360 degrees, in order to map out possible sources of systematic errors due to inhomogeneity.

{\bf \paragraph{HLSS Relative Flux Calibration} The relative spectrophotometric flux calibration shall be known to 2 percent relative accuracy (with the goal of 1\%), in order to understand the effective sensitivity limit for each redshift bin for each area surveyed.}

The requirement here is only on the {\it relative} spectrophotometry, which impacts the selection function of galaxies. Absolute line flux calibration will only change the overall number of objects and the d$N/$d$z$, but will not introduce density variations.  Large scale structure measurements require precise knowledge of the selection function of galaxies. Although the overall redshift distribution may be determined by averaging over the entire survey, fluctuations in the selection function can easily contaminate the underlying cosmological density fluctuations.

The HLSS spectroscopic sample is expected to be defined by a line flux
limit of 10$^{-16}\,$erg$\,$s$^{-1}$cm$^{-2}$. Spatial errors in the spectrophotometric calibration will introduce artificial spatial fluctuations in the number density of galaxies, which could contaminate the cosmological signal. To control this systematic effect, we require that the non-cosmological fluctuations in the mean number density (or the selection function of the survey) be $< 1\%$ (sqrt variance) when averaged over spatial scales between 10 Mpc/$h$ to 200 Mpc/$h$. At small scales, this is $\sim$ two orders of magnitude smaller than the cosmological signal, while at the $\sim$ BAO scale of 100 Mpc/$h$, this is $\sim$ one order of magnitude smaller than the cosmological signal.  These fluctuations equal the cosmological signal at $\sim400 \mathrm{Mpc}/h$.  These physical scales correspond to $\sim0.5$ degrees to 6 degrees at a redshift of 1.5.

The above requirement can be converted to a requirement on the spectrophotometric calibration accuracy, giving the luminosity function of galaxies \citep{Pozzetti:2016,Zhai_2019}. At the line flux limit of \romanM, this yields a requirement of $\sim$1\% relative spectrophotometric calibration, averaged over angular scales of 0.5 degrees to 6 degrees.
This is a very stringent requirement. We have relaxed this requirement from 1\%
to 2\% to add margin for mission success, assuming that we will achieve 1\%
relative spectrophotometric flux calibration in post-processing by projecting
out problematic modes in the analysis, and the use of Ubercal to minimize zero-point flux fluctuations \citep{Pad08,Markovic16}.

{\bf \paragraph{HLSS Data Processing} The data processing system shall provide sufficient knowledge of the 3D selection function so that the artificial correlations due to inaccuracies in the 3D selection function are less than 10\% of the statistical error bars on scales smaller than 2 degrees, and less than 20\% on larger angular scales.}

This requirement is on the contribution to the total error budget by the uncertainties in the 3D selection function. The BAO scale is less than $\sim$ 2.6 degrees (assuming Planck 2018 cosmology) in the redshift range for the HLSS ($1<z<3$). To convert the positions of observed galaxies in the large-scale structure into clustering measurements (correlation function, power spectrum, higher order statistics) we need to know how the ``average'' number density of objects (in the absence of clustering) changes in the observed volume. The mean number density will vary significantly both in redshift and with angular position due to effects of target selection, data reduction and observing conditions. Previous surveys were able to separate the selection function into two independent parts: the radial selection function and the angular selection function. It is likely that the \romanM\ selection function will not be separable in this way, i.e. different parts of the sky will have different radial profiles. 
Note, however, most effects are either mostly radial (e.g. target selection, data reduction) or angular (e.g. imaging quality, Galactic extinction).

The knowledge about 3D selection function is usually encoded into sets of random catalogs. When computing clustering statistics, the random catalogs remove the systematic effects of varying mean number density (due to target selection, data reduction or observing conditions). If the 3D selection function is not correct, the effects will not be completely removed and will generate spurious correlations that can bias the true cosmological signal. The angular mask of the \romanM\ HLSS will vary pixel to pixel on the NIR detector. The full description of the angular mask may turn out to be computationally intractable. For the core science goals we require the description of the mask to be correct with an angular resolution of approximately 3 arcmin. This corresponds to a spatial resolution of $3 \,h^{-1} \mathrm{Mpc}$ at $z=1.5$. This is driven by the fact that we need to be able to resolve the BAO peak. In principle, our requirements on the knowledge of the 3D selection function are driven by the main requirement that the spurious correlations should be no more than 10\% of statistical errors between the scales of 10 and 150 h$^{-1}$Mpc in clustering signals (either in correlation function multipoles or power-spectrum).

\subsection{Survey design}
\label{sec:ref-survey}
The main driver for the survey area of the \romanM\ HLSS is to maximize overlap with the \romanM\ HLIS, to use the HLIS for spectroscopic source identification and building the template model of contaminating background for each source using multi-band photometry.
The main driver for the depth of the \romanM\ HLSS is to go significantly deeper than \euclid, for robust modeling of systematic uncertainties, and increasing the power of higher-order statistics as a dark energy probe.

The observing plan for the HLSS has several free parameters -- the area vs.\ depth trade, the location of the footprint on the sky, the tiling/dithering pattern, and the timing of the observations. All of these are subject to trade, and will be re-optimized based on the science landscape prior to \romanM\ launch, but for the purposes of writing \romanM\ requirements and showing that they can be met in a 5-year mission, we have designed a Reference Survey.

\begin{figure}
\begin{center}
\includegraphics[width=6.5in]{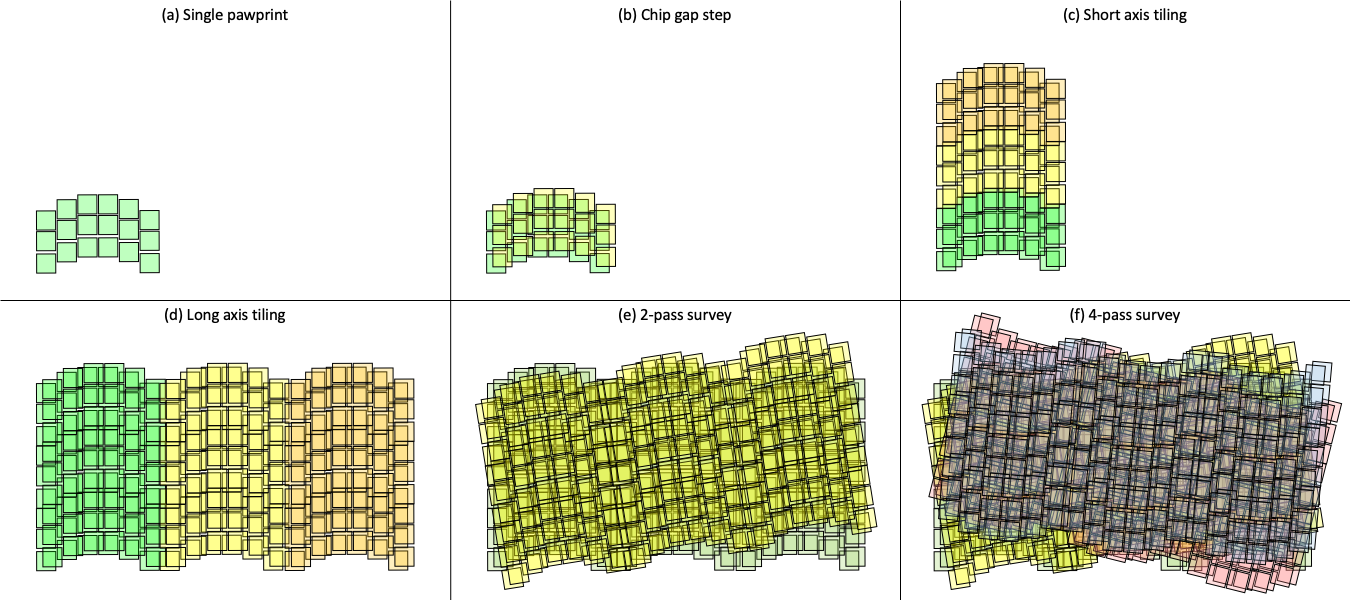}
\end{center}
\caption{\label{fig:tile}A cartoon of the HLSS tiling pattern. Panels (a--f) show the sequence from a single pointing to successively larger regions.}
\end{figure}

The tiling pattern for the Reference Survey is built hierarchically, starting from a single 0.28 deg$^2$ pointing and then building a nested series of {\tt for} loops until we reach the whole survey with multiple exposures and roll angles, as shown schematically in Figure~\ref{fig:tile}. The pattern itself is constrained by the layout of the \romanM\ focal plane. \romanM\ is a three-mirror anastigmat optical design, which can correct an annular field or portion thereof \citep[e.g.][]{1977ApOpt..16.2074K}; the wide-field instrument contains 18 detectors using a portion of this annulus \citep{2018SPIE10745E..0KP}. These are arranged in a $6\times 3$ pattern, but with an arc to fit all detectors within the annulus (see Fig.~\ref{fig:tile}a). The detectors each map to a 7.5$\times$7.5 arcmin region on the sky, but have gaps of 0.6--1.5 arcmin; to mostly cover these gaps, we do a diagonal chip gap dither and take a second exposure (Fig.~\ref{fig:tile}b).

The next level of the observing pattern hierarchy is to tile the sky, since \romanM\ has only one grism (unlike \euclid) and it takes much longer to roll to a new orientation than to slew to an adjacent field. We tile first along the short axis (Fig.~\ref{fig:tile}c) to make a strip, and then lay strips next to each other to make a 2D tiling (Fig.~\ref{fig:tile}d). The short axis was placed in the inner {\tt for} loop both for efficiency (the 0.4 deg short-axis slews are faster than the 0.8 deg long-axis slews), and because we found in observing simulations that strips with smooth edges can be tiled on a curved sky whereas the ``spaghetti strips'' that form if the long axis is in the inner {\tt for} loop suffer severe inefficiencies.

Finally, we repeat the pattern first at a slightly offset roll angle (Fig.~\ref{fig:tile}e) and then do 2 more rolls with the observatory rolled almost 180$^\circ$ (Fig.~\ref{fig:tile}f). Even more rolls are observed in the \romanM\ Deep Fields (required for calibration). These rolls both reduce spectral confusion, and -- by providing many repeated observations of the same stars at different points on the focal plane -- are expected to improve the ubercalibration solution \citep{Markovic16}. In the absence of chip gaps, this would lead to a total of 8 observations (2 each at 4 roll angles), but most points in the survey area fall into a gap at least once. Simulations of the observing plan give a coverage of 487, 1162, 1712, 1961, 2046, 2076, 2102, and 2117 deg$^2$ with coverage of $\ge 8$, 7, 6, 5, 4, 3, 2, and 1 grism observation, respectively. The cases with $<4$ observations are typically due to edge effects where only some roll angles are observed, and are unlikely to be used for the science analysis. Each observation is $\sim$ 300s in duration in the Reference HLSS.

The Reference HLSS requires 214 days (including observing overheads and deep fields for HLSS calibration, but not including additional calibration observations that are budgeted separately). It is not conducted in one single chunk, but will be distributed throughout the \romanM\ mission, constrained by (i) the field of regard (\romanM\ observes between 54 and 126 degrees from the Sun), (ii) the roll constraints (\romanM\ can roll $\pm 15$ degrees from the optimal angle on the solar panels), and (iii) the other observing programs. In particular, we do not do HLSS observations during the microlensing program (which takes 6 72-day ``seasons'' centered in March and September when the Galactic Bulge is available), and the HLIS and HLSS must be interlaced with the supernova program (which requires regular returns to the supernova field over the course of 2 years, but at only 25\%\ duty cycle). The roll constraint implies that observations where the observatory rolls by $\approx 180^\circ$ must be carried out with a separation of 6 months (plus an integer number of years), where the Sun is on the opposite side of the observatory. Fig.~1 of \cite{Eifler21a} shows one possible example of the scheduling.

\begin{figure}
\includegraphics[width = 3.5in]{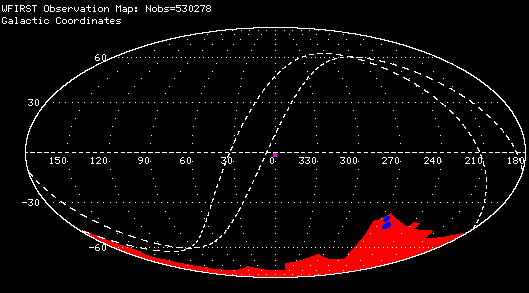}
\includegraphics[width = 3.5in]{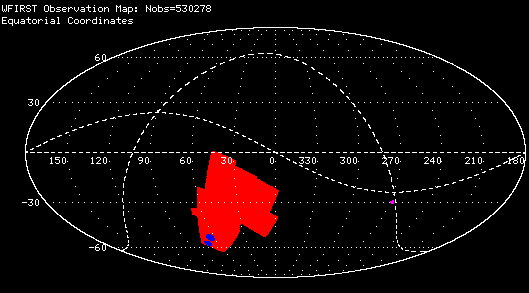}
\label{fig:footprint}
\caption{Notional footprint of the \romanM\ HLSS, in Galactic (left) and Equatorial (right) coordinates. The small blue regions indicate notional \romanM\ HLSS Deep Fields for calibration; they are placed within the \romanM\ continuous viewing zone to maximize efficiency of the observations.
}\end{figure}

Fig.~\ref{fig:footprint} shows the notional footprint of the HLSS in the sky (Galactic and Equatorial views). 
The survey footprint for the HLIS + HLSS is placed at high Galactic latitude (to reduce dust extinction and confusion from stars) and at high Ecliptic latitude (to reduce zodiacal light; this also expands the range of roll angles allowed, as we discuss below). Since the Galactic and Ecliptic planes are tilted $60^\circ$ to each other, this leads to two large, contiguous regions on the sky where we could place the \romanM\ footprint: a ``Southern'' region and a ``Northern'' region. The Reference HLIS + HLSS was placed in the Southern region because the photometric redshifts for the weak lensing analysis require both \romanM\ near-infrared photometry and deep optical photometry, which is likely to come from the Vera Rubin Observatory based in the Southern Hemisphere.\footnote{https://www.lsst.org/} A portion of this footprint can extend up to $\approx 0^\circ$ declination because the Ecliptic is tilted 23$^\circ$ to the celestial Equator; we selected this part of the Southern region so that part of the \romanM\ footprint could be observed with Northern Hemisphere telescopes. An alternative to the Reference design would be to move part of (or extend) the \romanM\ footprint to the Northern region and cover it with optical imaging with Hyper Suprime Cam.\footnote{https://www.naoj.org/Projects/HSC/}

Finally, we come to the choice of roll angles in the survey. While we plan to simulate several choices, the range of possibilities is restricted by observing geometry, especially at low Ecliptic latitude. The ``easiest'' roll angles to schedule (in the sense of passing all the constraints for the greatest number of days during the mission) occur when the Sun is 90$^\circ$ from the target and the solar arrays are oriented directly toward the Sun. Because of the clocking angle of the wide-field instrument relative to the solar array, this corresponds to dispersion direction (vertical direction in Fig.~\ref{fig:tile}a) at Ecliptic position angle of $30^\circ$ or $210^\circ$. Using spherical trigonometry, one can show that the possible dispersion directions come in two ``fans'' at $30^\circ\pm\Delta\psi$ and $210^\circ\pm\Delta\psi$, where
\begin{equation}
\Delta\psi = 15^\circ + \sin^{-1}\left( \tan 36^\circ\,\tan|\beta|\right),
\end{equation}
where $\beta$ is the Ecliptic latitude of the target; $15^\circ$ is the maximum roll of the solar array relative to the Sun; and we recall that the allowed range of Sun-to-target angle is $90^\circ\pm36^\circ$. Thus we have $\Delta\psi = 15^\circ$ on the Ecliptic, rising to 26$^\circ$ at $|\beta|=15^\circ$; 39$^\circ$ at $|\beta|=30^\circ$; and 62$^\circ$ at $|\beta|=45^\circ$. At Ecliptic latitudes $|\beta|>53^\circ$ (i.e., $\Delta\psi>90^\circ$), the two allowed ranges merge and all roll angles become possible.

\section{Simulations of Astrophysical Data}

To simulate the \romanM\ galaxies, we use the semi-analytic model (SAM) of galaxy formation, \glc\ \citep{benson_galacticus:_2012}, to add galaxies to cosmological dark matter only N-body simulations. This allows us to forecast the 3D distribution and properties of galaxies in the \romanM\ HLSS. Appendix A describes the implementation of \glc\ in detail. 

\subsection{Predicting emission line galaxy number densities}
\label{sec:ngal}

Galaxy number density as a function of redshift is the most important ingredient in investigating the constraining power of future galaxy redshift surveys and optimizing their survey designs. With \glc, we can perform realistic simulations to produce synthetic galaxy catalogs and predict the characteristics of the galaxy distributions expected from \romanM. The \glc\ model can output a set of galaxy properties, such as redshift, stellar mass, star formation rate, metallicity, photometric luminosities for a set of filter transmission curves. In addition, \glc\ can be coupled with \cloudy\ photoionization code (\citealt{ferland_2013_2013}) to compute the emission line luminosity, see \S\ref{galacticus:differential} (see also \citealt{Merson_2018}). The basic idea is to interpolate the tabulated libraries of emission line luminosities using \glc\ output for each galaxy, including the ionizing photons, metallicity of the interstellar medium, hydrogen gas density and so on. Due to the coupled nature of multiple baryonic processes and our poor prior knowledge, the model needs calibration to make reasonable predictions.

Since \romanM\ HLSS will target emission line galaxies at redshift $z>1$, we calibrate the \glc\ model using observations at higher redshift, instead of focusing on the local Universe. In particular, we use the measurements of WISP number counts for H$\alpha$ emitters at $0.7<z<1.5$ (\citealt{Mehta_2015}), and the H$\alpha$ luminosity function from 
the ground-based narrow-band High-z Emission Line Survey (HiZELS, \citealt{Geach_2008, Sobral_2009, Sobral_2013}), at $z=0.4, 0.84, 1.47$ and $2.23$. 
In addition to the parameters that describe the physics governing galaxy formation, we employ the \cite{Calzetti_2000} model for dust attenuation and tune the strength of attenuation by varying the parameter $A_{V}$. The value of this parameter can impact the overall luminosity of the nebula emission of galaxies. 

The merger trees of dark matter halos are the input to the \glc\ SAM, and are constructed from the UNIT simulation (\citealt{Chuang_2019}), which assumes a spatially flat $\Lambda$CDM model with 
$\Omega_m = 0.3089$, $h \equiv H_0/100 = 0.6774$, $n_s= 0.9667$, and $\sigma_8 = 0.814$ (see Table 4 in \citealt{Planck_2016}).
The UNIT simulation has a volume of 1($h^{-1}$Gpc)$^{3}$ with a mass resolution of $\sim10^{9}h^{-1} M_{\odot}$. This enables the identification of subhalos over a sufficient cosmic volume for statistics of both dark matter halos and galaxies. In order to forecast the number density of ELGs from \romanM\ HLSS, we run the \glc\ model on top of the UNIT simulation to build a lightcone catalog, covering an area of 4 deg$^{2}$ in the redshift range of $0<z<3$. Our \glc\ model constructs the lightcone mock using the methodology presented by \cite{Kitzbichler_2007} which identifies dark matter halos that intersect the lightcone of the observer. We utilize all 129 snapshots from the UNIT simulation back to $z=3$ in constructing this lightcone, but do not perform any interpolation of galaxy positions or other properties between snapshots to the precise redshift at which the galaxy intersects the lightcone. The snapshots are close enough in separation that interpolation is not required.
The model also replicates the simulation box to cover sufficient volume for our mock. The method of \cite{Kitzbichler_2007} chooses the line of sight through the simulation to minimize any self-intersection of the lightcone in the periodic replications of the simulation cube. Our lightcone has a cross section of 154 by 154 comoving $h^{-1}$ Mpc at $z=3$, which is much smaller than the UNIT simulation cube which is $1 h^{-1}$Gpc (comoving) on a side. As such, self-intersection and any repeated structure in the lightcone will be minimal.
Although our simulation is carried out for simulating the \romanM\ HLSS, the data products can have much broader use, e.g., application to surveys such as the one planned for \euclid, or the investigation of ELG properties compared with current observations. This simulation also outputs the spectral energy distribution (SED) of galaxies within the wavelength range of the \romanM\ grism, as input to the grism simulations for \romanM\ HLSS (see \S\ref{sec:grism_sim}).

\begin{figure}
\includegraphics[width = 3.5in]{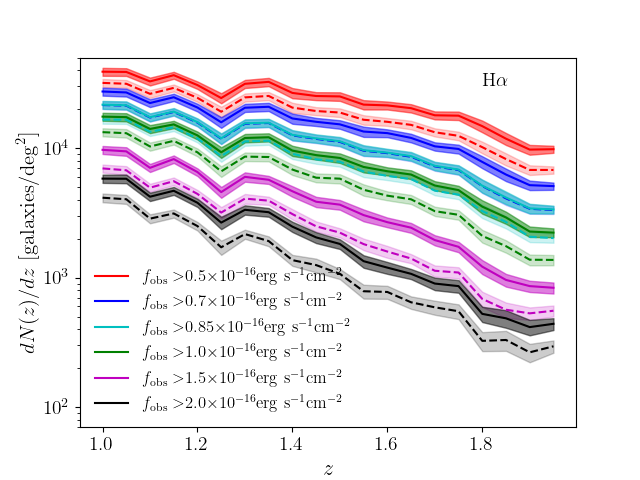}
\includegraphics[width = 3.5in]{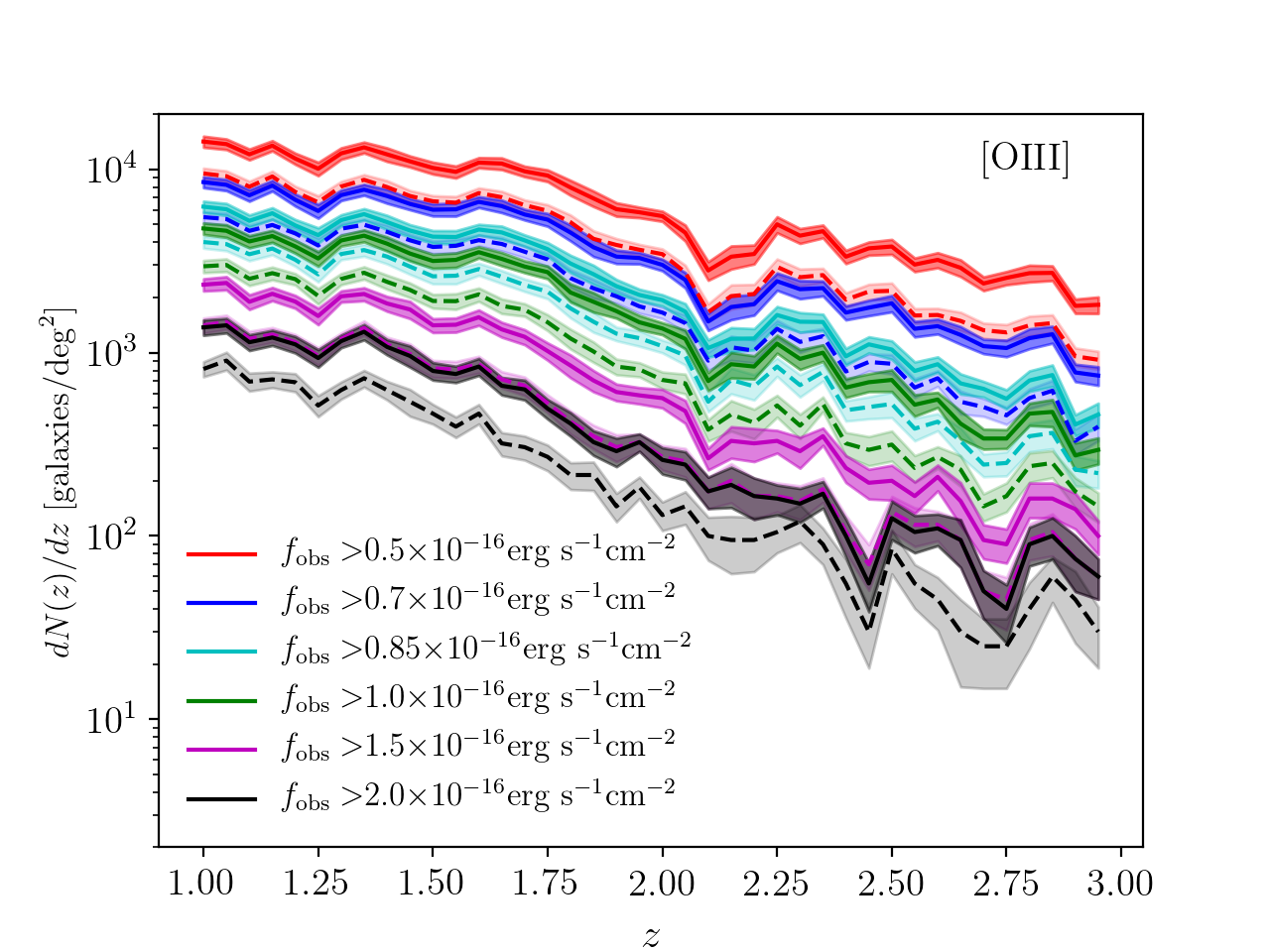}
\label{fig:ELG_number_density}
\caption{Number density of ELGs as a function of redshift for H$\alpha$ (left) and [OIII] (right). The result shows the effect from dust attenuation model and emission line flux. The errors are a combination of jackknife uncertainty and cosmic variance, shown as the shaded area. The solid lines adopt dust attenuation parameter $A_{V}=1.65$, while the dashed lines are for $A_{V}=1.92$.}
\end{figure}

Due to the uncertainties in the dust attenuation model, we present the galaxy mock with two different values of $A_{V}$. The predicted number density of \romanM\ ELGs as a function of redshift is shown in Figure \ref{fig:ELG_number_density} for both H$\alpha$ and [OIII] emitters, and is also summarized in Tables \ref{tab:Halpha} and \ref{tab:OIII}. The dust models have $A_{V}=1.652$ and $A_{V}=1.915$ respectively, they correspond to calibration of \glc\ to match the WISP number counts for H$\alpha$ emitters (\citealt{Mehta_2015}) at redshift $0.7<z<1.5$, and H$\alpha$ luminosity function from HiZELS \citep{Sobral_2013}. We have investigated the impact of the survey depth on the number density estimates by considering six thresholds in our analysis: $2.0\times10^{-16}\text{erg/s/cm}^{2}$ (the $3.5\sigma$ depth of a \euclid-like galaxy redshift survey), $1.0\times10^{-16}\text{erg/s/cm}^{2}$ (the $6.5\sigma$ depth of the Reference \romanM\ HLSS), $8.5\times10^{-17}\text{erg/s/cm}^{2}$ (the $5\sigma$ depth of the Reference \romanM\ HLSS), $1.5\times10^{-16}\text{erg/s/cm}^{2}$, $7\times10^{-17}\text{erg/s/cm}^{2}$, and $5\times10^{-17}\text{erg/s/cm}^{2}$. The latter three are included to facilitate a depth versus area optimization of \romanM\ HLSS.
The predicted galaxy number counts are sensitive to the line flux limit, as expected. 
The Reference \romanM\ HLSS survey provides an improvement by a factor of a few compared with \euclid\ in survey depth for H$\alpha$ ELGs, and enables observations of a sufficiently large number of [OIII] ELGs at higher redshifts for BAO/RSD measurements. For \romanM\ HLSS, we can expect thousands of H$\alpha$ emitters per square degree in the redshift range $1<z<2$. This number density is critical for the clustering analysis to retrieve cosmological information without being dominated by shot noise.

Although we have calibrated \glc\ SAM using data from H$\alpha$ observations only, its predictions for [OIII] ELGs are consistent with available observational data (see more details in \citealt{Zhai_2019}). The right hand panel of Figure \ref{fig:ELG_number_density} shows the predicted number density of [OIII] ELGs as a function of redshift for different dust models and flux limits. At redshift $z<2$, the [OIII] emission line can be used along with the H$\alpha$ line (the primary line) for robust redshift determination for galaxies from the \romanM\ HLSS. At redshift $z>2$, [OIII] becomes the primary line of ELGs, and [OII] becomes the secondary line, in determining redshifts for galaxies from the \romanM\ HLSS. At the nominal depth of $1.0\times10^{-16}\text{erg/s/cm}^{2}$ ($6.5\sigma$), the Reference \romanM\ HLSS can observe hundreds of [OIII] emitters per square degree within $2<z<3$. Although this number density is degraded by roughly an order of magnitude compared with the H$\alpha$ counterparts at lower redshifts, the [OIII] ELGs are highly biased tracers of large scale structure at $2<z<3$, which partially offsets the decrease in number density $n(z)$ since the shot noise of the BAO/RSD measurements depend on $n(z) b(z) P_m(k,z)$ (where $b(z)$ and $P_m(k,z)$ are the galaxy bias and matter power spectrum respectively).
Thus the \romanM\ [OIII] ELGs will be a useful cosmological probe out to a redshift $\sim$3,  enabling the measurement of cosmic expansion rate and growth history of structure when the universe is only $\sim2$ Gyrs old.

\begin{sidewaystable}
\centering
\small\addtolength{\tabcolsep}{-4pt}
\caption{Number counts of H$\alpha$-emitting galaxies, ${\rm d}N/{\rm d}z$, per square degree as a function of redshift for different line flux limits, predicted from Galacticus simulation. The uncertainties are estimated from jackknife resampling method and cosmic variance. The results for both of the two dust models are shown. Note that the observational efficiency is not included here.}
\begin{tabular}{cccccccc|cccccc}
\hline
\multicolumn{2}{c}{ }  &  \multicolumn{5}{c}{Dust model with $A_{V}=1.65$} & \multicolumn{5}{c}{Dust model with $A_{V}=1.92$} \\
\hline
\multicolumn{2}{c}{Redshift} & \multicolumn{5}{c}{Flux limit [erg s$^{-1}$cm$^{-2}$]} & \multicolumn{5}{c}{Flux limit [erg s$^{-1}$cm$^{-2}$]}\\
From & To & $\left(5\times 10^{-17}\right)$ &  $\left(7\times 10^{-17}\right)$ &  $\left(8.5\times 10^{-17}\right)$ & $\left(1\times 10^{-16}\right)$ &  $\left(1.5\times 10^{-16}\right)$ &$\left(2\times 10^{-16}\right)$ & $\left(5\times 10^{-17}\right)$ &  $\left(7\times 10^{-17}\right)$  &  $\left(8.5\times 10^{-17}\right)$ & $\left(1\times 10^{-16}\right)$ & $\left(1.5\times 10^{-16}\right)$   &$\left(2\times 10^{-16}\right)$ \\
\hline
0.5 &  0.6  & $34725\pm2775$  & $26812\pm2109$ &  $22912\pm1805$  & $19912\pm1550$  &  $13662\pm1070$  & $9820\pm775$    & $29787\pm2363$  & $22795\pm1800$ & $19327\pm1506$ & $16642\pm1311$  & $10935\pm863$   &  $7732\pm610$      \\
0.6 &  0.7  & $42560\pm2922$  & $32052\pm2197$ &  $26952\pm1878$    & $23110\pm1600$  & $14995\pm1057$ & $10420\pm770$    & $36100\pm2481$  & $26830\pm1868$ & $22317\pm1548$ &  $18820\pm1308$  &  $11657\pm839$   &  $7820\pm591$      \\
0.7 &  0.8  & $34850\pm2025$  & $26302\pm1527$ &  $22022\pm1285$ & $18567\pm1088$   &  $11675\pm686$  & $7802\pm470$     & $29610\pm1734$  & $21915\pm1280$  & $17890\pm1043$ &  $14985\pm879$    &  $8875\pm541$  &  $5815\pm354$      \\
0.8 &  0.9  & $35750\pm2205$  & $26277\pm1647$ &  $21670\pm1380$ & $17942\pm1132$   &  $10555\pm670$   & $6897\pm459$     & $29992\pm1859$  & $21562\pm1374$ &   $17205\pm1083$  &  $14072\pm874$   &  $7952\pm511$  &  $4920\pm341$      \\
0.9 &  1.0  & $32662\pm1836$ & $23500\pm1335$  &  $18935\pm1081$ &$15570\pm892$     &   $8752\pm507$  & $5527\pm340$       & $27007\pm1531$  & $18835\pm1074$ & $14780\pm846$ & $11812\pm673$   &  $6445\pm401$  &  $3960\pm261$      \\
1.0 &  1.1  & $36517\pm2225$ & $25317\pm1575$  &  $20057\pm1238$ & $16235\pm1013$   & $8780\pm549$   & $5340\pm337$    & $29597\pm1828$  & $19932\pm1229$  & $15470\pm956$ & $12120\pm759$  &  $6282\pm395$  &  $3667\pm242$       \\
1.1 &  1.2  & $34552\pm1983$ & $23292\pm1372$  &  $17970\pm1075$  & $14312\pm847$    &  $7422\pm444$    & $4260\pm266$    & $27627\pm1595$  & $17837\pm1069$ & $13595\pm805$ &  $10620\pm622$  &  $5042\pm304$  &  $2865\pm182$      \\
1.2 &  1.3   & $26615\pm2061$  & $17502\pm1394$ & $13287\pm1060$ & $10377\pm840$   &  $5280\pm433$   & $3055\pm262$     & $20987\pm1646$  & $13202\pm1051$ &  $9777\pm794$  & $7537\pm607$    &  $3625\pm301$  &  $1990\pm172$      \\
1.3 &  1.4  & $31170\pm2009$  & $20030\pm1288$ & $15107\pm949$  & $11615\pm740$  &   $5552\pm350$  & $3025\pm200$     & $24215\pm1550$  & $14995\pm944$ & $10900\pm693$  & $8217\pm521$    &   $3727\pm245$   &  $1835\pm128$       \\
1.4 &  1.5  & $26380\pm1708$  & $16642\pm1078$ & $12372\pm831 $ & $9270\pm632$    &  $4267\pm310$  & $2247\pm172$    & $20185\pm1309$  & $12272\pm823$ &  $8692\pm598$  &  $6410\pm454$   &  $2765\pm204$   &  $1327\pm115$        \\
1.5 &  1.6  & $22125\pm1442$    & $13555\pm934$ & $9770\pm688$  & $7190\pm505$    &  $2985\pm219$   & $1355\pm100$     & $16742\pm1116$  & $9675\pm679$ & $6625\pm460$  & $4790\pm345$  &  $1785\pm127$    &  $800\pm59$       \\
1.6 &  1.7  & $19727\pm1144$   &  $11752\pm689$ & $8432\pm503$ & $6075\pm373$   &  $2342\pm159$  & $1050\pm78$     & $14670\pm869$   & $8350\pm497$ & $5582\pm346$ & $3887\pm254$    & $1352\pm94$   &  $667\pm56$       \\
1.7 &  1.8  & $17817\pm1174$   & $9895\pm680$  & $6795\pm477$  & $4705\pm322$   &  $1740\pm135$   & $832\pm70$    & $12645\pm848$   & $6732\pm472$ &  $4345\pm298$ & $2997\pm212$   & $1072\pm82$  &  $542\pm54$       \\
1.8 &  1.9   & $11732\pm957$  & $6157\pm505$ &  $4092\pm343$ & $2792\pm233$    &  $955\pm83$    & $477\pm51$    & $8095\pm661$  & $4047\pm339$ & $2562\pm220$ & $1640\pm140$   &  $575\pm56$  &  $310\pm34$      \\
1.9 &  2.0   & $9955\pm547$  & $5070\pm285$  & $3340\pm199$ &  $2232\pm139$   &  $820\pm58$    & $390\pm32$    & $6792\pm388$  & $3305\pm196$ & $2027\pm127$ & $1367\pm96$   & $502\pm38$   &  $255\pm22$      \\
\hline
1.0 &  2.0  & $23659\pm1201$  & $14921\pm762$ & $11122\pm569$ & $8480\pm432$   &  $4014\pm206$   & $2203\pm114$    & $18155\pm924$  & $11035\pm564$ & $7957\pm405$ & $5958\pm303$  & $2673\pm138$   & $1426\pm75$ \\
0.5 &  2.0  & $27809\pm1404$ & $18944\pm958$  & $14914\pm758$  & $11993\pm610$  &  $6652\pm341$   & $4166\pm217$   & $22270\pm1125$ & $14819\pm753$ & $11406\pm580$ & $9061\pm461$  & $4839\pm250$  & $2967\pm156$   \\
\hline
\end{tabular}
\label{tab:Halpha}
\end{sidewaystable}

\begin{sidewaystable*}
\centering
\small\addtolength{\tabcolsep}{-2pt}
\caption{Same as Table \ref{tab:Halpha} but for [OIII]-emitting galaxies.}
\begin{tabular}{cccccccc|cccccc}
\hline
\multicolumn{2}{c}{ }  &  \multicolumn{5}{c}{Dust model with  $A_{V}=1.65$} & \multicolumn{5}{c}{Dust model with  $A_{V}=1.92$} \\
\hline
\multicolumn{2}{c}{Redshift} & \multicolumn{5}{c}{Flux limit [erg s$^{-1}$cm$^{-2}$]} & \multicolumn{5}{c}{Flux limit [erg s$^{-1}$cm$^{-2}$]}\\
From & To & $\left(5\times 10^{-17}\right)$ & $\left(7\times 10^{-17}\right)$ & $\left(8.5\times 10^{-17}\right)$ & $\left(1\times 10^{-16}\right)$   & $\left(1.5\times 10^{-16}\right)$   &$\left(2\times 10^{-16}\right)$ & $\left(5\times 10^{-17}\right)$ & $\left(7\times 10^{-17}\right)$  & $\left(8.5\times 10^{-17}\right)$ & $\left(1\times 10^{-16}\right)$   & $\left(1.5\times 10^{-16}\right)$  &$\left(2\times 10^{-16}\right)$ \\
\hline
1.0 &  1.1  & $10437\pm617$  & $6160\pm381$ & $4447\pm287$ & $3440\pm230$   & $1610\pm117$   & $977\pm74$     &  $6880\pm426$  & $3920\pm259$ & $2892\pm197$ & $2117\pm140$   &  $1032\pm76$   & $610\pm58$        \\
1.1 &  1.2  & $10212\pm589$ & $5907\pm364$ & $4225\pm257$ & $3167\pm204$    &    $1550\pm114$  & $922\pm72$    &  $6590\pm400$  & $3707\pm228$ & $2657\pm170$ & $1942\pm131$   &  $957\pm76$   & $557\pm47$         \\
1.2 &  1.3  & $8360\pm609$   & $4797\pm350$ & $3380\pm254$ & $2560\pm191$    &   $1155\pm101$ & $697\pm67$     &  $5362\pm387$  & $2992\pm222$ & $2145\pm164$ & $1562\pm132$   &  $727\pm69$   & $427\pm48$        \\
1.3 &  1.4  & $10027\pm616$ & $5727\pm344$ & $4125\pm246$ &$3095\pm189$   &   $1540\pm105$  & $875\pm64$     &  $6442\pm388$  & $3627\pm218$ & $2610\pm160$ & $1967\pm128$  &  $935\pm67$   & $572\pm45$        \\
1.4 &  1.5  & $8987\pm578$   &  $5147\pm361$ & $3690\pm275$ & $2732\pm196$    &    $1292\pm109$   & $752\pm67$   &  $5752\pm394$  & $3215\pm244$ & $2267\pm166$ & $1677\pm139$    &  $790\pm71$  & $482\pm49$        \\
1.5 &  1.6  & $7900\pm484$   & $4530\pm288$  & $3217\pm209$ & $2355\pm163$   &  $1015\pm77$  & $562\pm50$   &  $5155\pm328$  & $2780\pm191$ & $1857\pm137$ & $1350\pm99$   &  $585\pm52$   & $340\pm33$        \\
1.6 &  1.7  & $8365\pm495$   &  $4792\pm295$ & $3267\pm209$ & $2355\pm150$     &  $962\pm75$  & $490\pm45$    &  $5377\pm335$  & $2840\pm177$ & $1872\pm126$ & $1315\pm93$    &  $505\pm46$   & $290\pm29$        \\
1.7 &  1.8  & $6957\pm460$   & $3870\pm254$  & $2690\pm182$ & $1930\pm134$   & $812\pm64$   & $455\pm44$    &  $4362\pm286$  & $2340\pm153$ & $1555\pm107$ &$1110\pm77$  &   $477\pm44$    & $310\pm37$        \\
1.8 &  1.9  & $5412\pm415$  &  $2857\pm236$  & $1937\pm169$ & $1305\pm122$   &  $480\pm48$   & $262\pm31$    &  $3252\pm266$   & $1640\pm147$ & $1050\pm97$ & $682\pm66$   &  $277\pm32$   & $167\pm22$        \\
1.9 &  2.0  & $4442\pm251$   &  $2277\pm141$ & $1532\pm103$ & $1052\pm68$   & $445\pm33$ & $275\pm28$      &  $2680\pm161$   & $1305\pm89$ & $845\pm53$ & $605\pm40$     &  $282\pm29$   & $180\pm17$        \\
2.0 &  2.1  & $3055\pm237$ & $1725\pm135$ & $1185\pm104$ & $875\pm85$    &  $430\pm51$ & $302\pm37$      &  $1962\pm152$   & $1047\pm99$  & $715\pm71$ &  $555\pm62$   &  $317\pm39$    &  $235\pm32$         \\
2.1 &  2.2  & $2265\pm243$   & $1215\pm138$ & $817\pm100$ & $597\pm69$   &  $282\pm39$ & $190\pm30$      &  $1422\pm160$    & $710\pm84$  & $490\pm58$ &  $345\pm45$     &   $200\pm31$ &  $137\pm23$        \\
2.2 &  2.3  & $3210\pm287$ &  $1642\pm151$  & $1102\pm116$ & $715\pm79$    &  $307\pm38$ & $227\pm31$      &  $1872\pm175$    & $937\pm99$  & $567\pm63$ &  $395\pm46$   &  $230\pm31$   &  $175\pm22$         \\
2.3 &  2.4   & $3170\pm230$  & $1615\pm129$ & $1097\pm93$& $820\pm69$    &   $440\pm37$   & $302\pm32$     &  $1845\pm137$    & $980\pm81$  & $687\pm56$  &  $522\pm41$    &  $310\pm34$ &  $265\pm26$        \\
2.4 &  2.5  & $2722\pm219$  & $1305\pm100$ & $817\pm71$ & $520\pm54$   & $212\pm30$ & $122\pm21$    &  $1547\pm114$    &  $667\pm61$  & $375\pm44$ &  $267\pm33$   &  $122\pm21$   &  $97\pm17$         \\
2.5 &  2.6  & $2360\pm169$  & $1087\pm88$  & $677\pm65$ & $477\pm50$   &  $237\pm27$  & $162\pm21$     &  $1297\pm103$    & $587\pm59$ & $377\pm42$ &  $275\pm32$   &  $175\pm23$  &  $120\pm19$         \\
2.6 &  2.7  & $2090\pm169$   & $850\pm76$ & $480\pm44$ & $312\pm33$    &  $117\pm22$   & $67\pm12$      &  $1067\pm93$   &  $395\pm43$ & $245\pm31$ &  $160\pm23$     &  $72\pm12$  &   $30\pm8$          \\
2.7 &  2.8  & $1932\pm149$   &  $795\pm67$ & $435\pm36$ & $295\pm33$   &  $105\pm14$  & $60\pm11$       &  $960\pm78$    & $372\pm34$  & $225\pm28$ &  $140\pm19$    &   $60\pm11$    &  $35\pm8$         \\
2.8 &  2.9  & $1942\pm187$  &  $847\pm95$  & $477\pm60$ & $337\pm46$  &  $155\pm23$   & $105\pm20$    &  $1045\pm106$  & $405\pm54$ & $265\pm39$  &  $195\pm29$    &  $110\pm20$  &  $67\pm11$        \\
2.9 &  3.0  & $1337\pm117$   &  $522\pm61$ & $345\pm40$ & $222\pm29$   &  $100\pm19$   & $67\pm15$   &  $625\pm66$    & $290\pm33$  & $167\pm22$ &  $122\pm18$     &  $70\pm14$  &  $35\pm11$        \\
\hline
1.0 &  2.0  & $8110\pm406$  &  $4606\pm230$  & $3251\pm164$  & $2399\pm121$   & $1086\pm54$  & $627\pm32$     & $5185\pm258$  & $2836\pm143$ & $1975\pm99$  &  $1433\pm72$   &  $657\pm33$  &  $393\pm20$      \\
2.0 &  3.0  &  $2408\pm127$  &   $1160\pm62$ & $743\pm42$ & $517\pm30$   & $238\pm15$  & $160\pm10$    & $1364\pm74$  &  $639\pm37$ & $411\pm24$ & $297\pm18$    &  $166\pm11$ &  $119\pm7$        \\
\hline
\end{tabular}
\label{tab:OIII}
\end{sidewaystable*}

\subsection{Galaxy mock catalog}
\label{sec:mock}

In order to facilitate the study of BAO/RSD and other cosmological measurements from \romanM\ HLSS, we have produced a mock data set, a lightcone catalog of galaxies with an area of 2000 deg$^{2}$ using the calibrated parameters for \glc\ and the dust model, consistent with the current baseline. This catalog is constructed using the same method as described in the previous section, but pays particular attention to the emission lines. In addition, it can also be used to study different galaxy statistics, void statistics, and higher order clustering signals to further explore the cosmological information from \romanM\ HLSS. 
To validate this mock galaxy catalog for the \romanM\ HLSS, we have performed a clustering analysis of H$\alpha$ ELGs from it, using techniques that have been used in analyzing actual data. Since this 2000 deg$^2$ mock is much larger than the previous one, the simulation needs multiple replications (3-4 replications in each direction transverse to the lightcone) of the UNIT box to give continuous distribution over the volume. This will lead to some repetition of structures in the lightcone on the largest scales. This can be improved with simulations of larger box size in the future.

For this mock data set, we adopt the dust model based on the measurements from HiZELS H$\alpha$ luminosity function, and select galaxies with H$\alpha$ flux higher than $1.0\times10^{-16}\text{erg/s/cm}^{2}$ within $1<z<2$. This results in a galaxy sample with $\sim10$ million H$\alpha$-emitting galaxies within a cosmic volume of 7$[\text{Gpc}/h]^{3}$. We then perform a likelihood analysis on this galaxy sample, measuring its galaxy clustering signal relative to a random catalog distributed within the same RA/DEC area, but with a number density ten times that of galaxies. 
The likelihood analysis compares the galaxy clustering measured from data to theoretical models in a Markov Chain Monte Carlo (MCMC) process \citep{Lewis_2002}, with a covariance matrix for the measurements derived from thousands of approximate mocks for galaxy clustering. 

We have adopted the EZmock method (\citealt{Chuang_2015}) to make the approximate mocks, and calibrate the parameters to have consistent clustering prediction as the fiducial model derived from the galaxy mock (the simulated data set). These mocks are constructed with a large enough cosmic volume and later truncated to have the same light cone specifications as the mock data set. The resultant EZmocks can mimic the fiducial model in terms of spatial distribution and clustering signal. This large number of mocks can provide accurate estimate of the covariance matrix. We analyze the clustering measurement in Fourier space, by comparing the power spectrum computed from the simulated galaxy data, and theoretical models. We refer the readers to \cite{Zhai_2021} for more details of the model and data analysis. 

In the MCMC likelihood analysis of the mock data set, we measure the key dark energy parameters from BAO/RSD: the scaled cosmic expansion rate $H(z)s$ and the scaled angular diameter distance $D_A(z)/s$ (both from BAO), and the scaled linear growth of large scale structure structure, $\overline{f_g(z) G(z)}$ (from RSD), where $s$ is the BAO scale (the comoving sound horizon at the drag epoch), $f_g(z)$ is the linear growth rate, and $G(z)$ is the linear growth factor.
We have marginalized the other parameters as nuisance parameters for this analysis. 
We find that the true values of the BAO/RSD parameters can be recovered within $\sim 1\sigma$ by our analysis. This verifies that the \romanM\ galaxy mock that we have created contains the correct information on galaxy clustering, and that information can be recovered by using the data analysis method currently used in analyzing actual data from galaxy redshift surveys.

\begin{figure*}
\centering
\includegraphics[width = 5in]{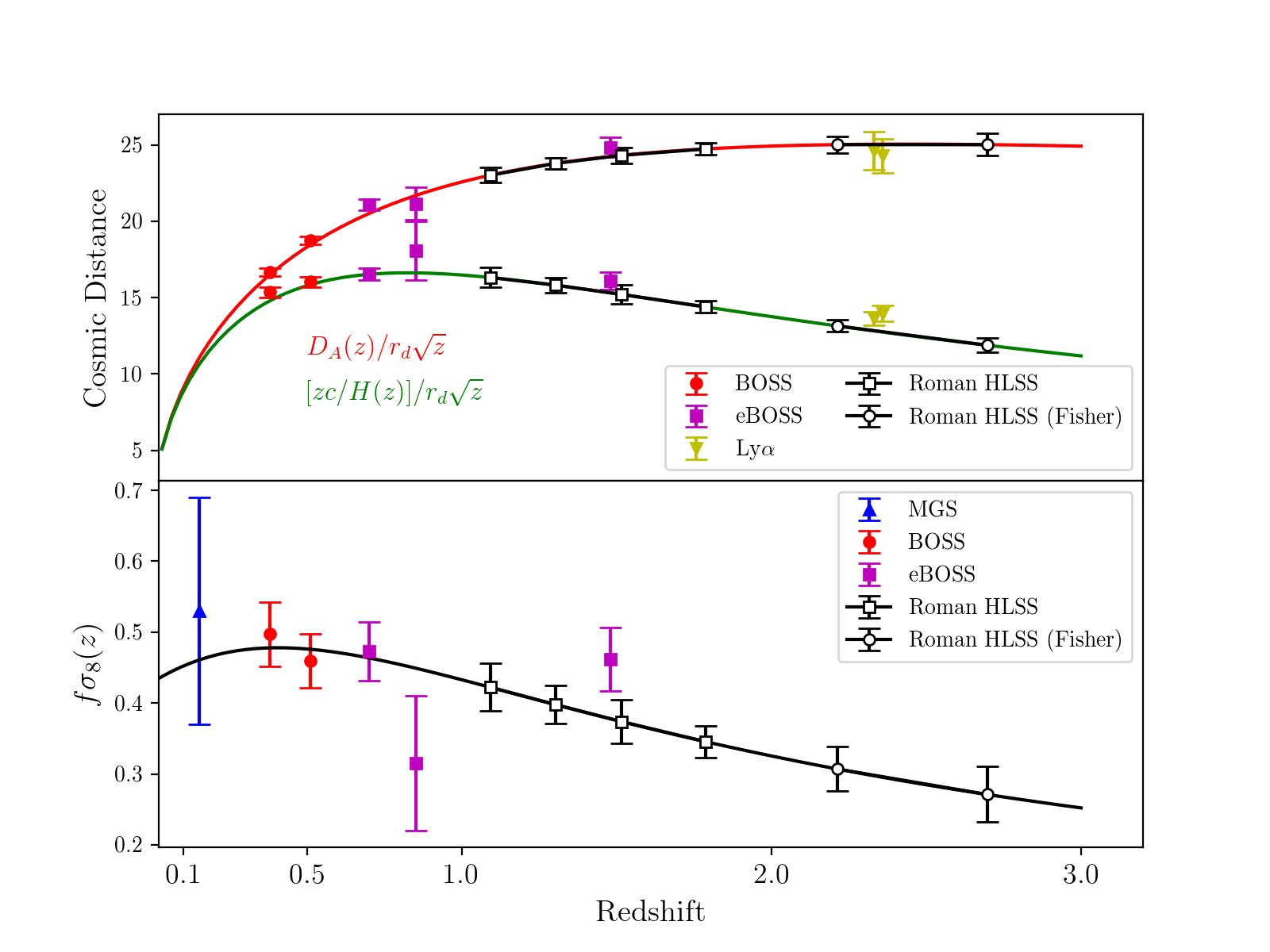}
\label{fig:BAO_RSD}
\caption{Cosmological measurements from the analysis of large scale structure in the Universe:  the cosmic distance scales from the BAO measurements (top panel), and the growth rate measurements (bottom panel). The current measurements compiled from \cite{eBOSS_2020} include observations from 6dFGS (\citealt{Beutler_2012}), SDSS MGS (\citealt{Ross_2015, Howlett_2015}), BOSS (\citealt{Alam_2017}), eBOSS (\citealt{Bautista_2020, GilMarin_2020, Tamone_2020, deMattia_2020, Hou_2020, Neveux_2020}) and BOSS-eBOSS Lyman-$\alpha$ forest (\citealt{Bourboux_2020}. The forecast for \romanM\ Reference HLSS are shown as empty points, based on the clustering analysis of H$\alpha$ ELGs at $1\leq z<2$ and [OIII] ELGs at $2\leq z<3$ respectively, assuming a Planck 2016 cosmology; these are very conservative by assuming S/N of 6.5 for the slitless spectra (see Sec.\ref{sec:fisher}). Note that the prediction of constraints for the Reference \romanM\ HLSS is from ``Model A" in \cite{Zhai_2021} at $1\leq z<2$, and from the Fisher matrix results presented in the left panel of Table \ref{tab:fisher1} at $2 \leq z <3$.}
\end{figure*}

Figure \ref{fig:BAO_RSD} shows the measurement errors from the \romanM\ HLSS mock data (assuming a Planck 2016 cosmology), compared with a compilation from current observations of large scale structure, including 6dFGS (\citealt{Beutler_2012}), SDSS MGS (\citealt{Ross_2015, Howlett_2015}), BOSS (\citealt{Alam_2017}), eBOSS (\citealt{Bautista_2020, GilMarin_2020, Tamone_2020, deMattia_2020, Hou_2020, Neveux_2020}) and BOSS-eBOSS Lyman-$\alpha$ forest (\citealt{Bourboux_2020}). Note that the analysis methods that we have used are not yet optimized, thus the forecast errors for the \romanM\ HLSS in Figure \ref{fig:BAO_RSD} are very conservative as a simple statistical analysis (see \S\ref{sec:fisher}). Future analysis should lead to higher precision, e.g., with improved modeling of non-linear scale information, enhanced BAO signal from reconstruction, and more robust estimate of the covariance matrix. 

Our \romanM\ HLSS galaxy mock catalog provides the first realistic simulation of the galaxy distribution in the \romanM\ HLSS. This data set can be used to study the optimization of constraints on the properties of dark energy and cosmological parameters using galaxy clustering statistics, and explore the systematic effects on the cosmological measurements. 

We have also created a 4 deg$^2$ galaxy mock with galaxy spectral energy distributions (SEDs) as input for the \romanM\ grism simulations (see Sec.\ref{sec:grism_sim}). We split the construction of SEDs into two components: continuum and emission lines. For the former, Galacticus tracks the star formation history of each galaxy on a grid of time and metallicity (optimized to provide high time resolution close to the time at which the galaxy is observed, and lower resolution further in the past). This star formation history is convolved with simple stellar population spectra from a stellar population library. This convolution process is carried out in the final production step such that all the millions of the galaxies in the catalog have a SED generated in a timely manner. For emission lines, we assume a Gaussian line profile, with a full-width-half-maximum determined by the rotation curve at the half-mass radius of the galaxy, normalized to the total line luminosity (see Sec.\ref{sec:ngal} and \citealt{Zhai_2019}). The final SED is the sum of the continuum and emission lines, coupled with the calibrated \cite{Calzetti_2000} dust model over the entire wavelength range. Our final SED data have a wavelength range of $0.2-4\mu$m and redshift range of $0<z<10$. This fully covers the \romanM\ HLSS requirement of $1.0-1.93\mu m$ as wavelength range within $0<z<3$. Depending on redshift, the resolution of the SED is R $\sim$ a few hundreds to 1000. In addition, the combination of this SED and any filter transmission curve can be used to compute broad-band photometry. These characteristics make the catalog useful not only for \romanM\ HLSS, but also applicable to other surveys, such as the \euclid\ surveys \citep{Euclid}, and the surveys proposed for ATLAS Probe \citep{Wang19}.

We note that measurements from our realistic \romanM\ galaxy mock indicate that the \romanM\ science requirements derived in \S\ref{subsec:sciencegoals} are fully met by the Reference HLSS.

\subsection{Linear bias of emission line galaxies}

With the large scale 2000 deg$^{2}$ galaxy mock for \romanM\ HLSS, we can perform a detailed clustering analysis under different selection conditions to forecast the linear bias of both H$\alpha$ and [OIII] ELGs. This is the key element for a follow-up Fisher matrix analysis to predict the constraining power of \romanM\ HLSS. 
The different emission lines have different wavelengths, thus correspond to different redshift ranges within the \romanM\ grism coverage of 1-1.93$\,\mu$m.
We focus on H$\alpha$ and [OIII] lines in the current analysis. Additional emission lines are present: [OII] as the 2nd line for [OIII] ELGs, [NII] and H$\beta$ as contaminants. These will be investigated in future work. Since H$\alpha$ emission line is detectable at $z<2.0$, we split the measurement of linear bias into two subsamples with $z<2.0$ and $z>2.0$.

\begin{figure*}
\centering
\includegraphics[width =7.0in]{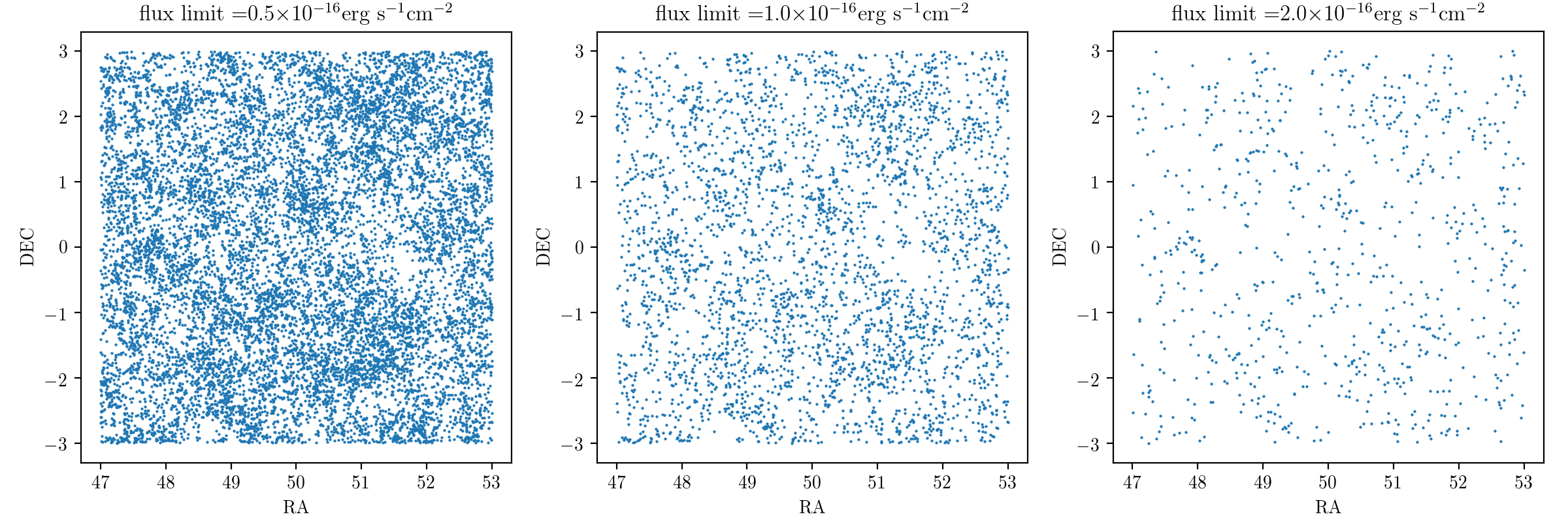}
\label{fig:radec}
\caption{The angular distribution of a subample of ELGs in the redshift range of $1.5<z<1.55$, for three different line flux limits:
$5\times10^{-17}\mathrm{erg}/\mathrm{s}/\mathrm{cm}^{2}$ (3D-HST 5$\sigma$ depth),
$10^{-16}\mathrm{erg}/\mathrm{s}/\mathrm{cm}^{2}$ (Reference \romanM\ HLSS 6.5$\sigma$ depth), and $2\times10^{-16}\mathrm{erg}/\mathrm{s}/\mathrm{cm}^{2}$ (Euclid 3.5$\sigma$ depth). The galaxies are chosen with both H$\alpha$ and [OIII] lines above the limit as indicated in each panel. The galaxy number density is very sensitive to the line flux limit.}
\end{figure*}

For galaxies with $z<2.0$, both H$\alpha$ and [OIII] are observed and can be used in the redshift determination. We apply flux limits to both two lines. The first limit $f_{\text{lim,1}}$ is the lower limit of the stronger line (either H$\alpha$ or [OIII]), chosen at $[0.5, 0.7, 1.0, 2.0]\times10^{-16}\mathrm{erg}/\mathrm{s}/\mathrm{cm}^{2}$. The second parameter $f_{\text{lim,2R}}$ sets the lower limit of the weaker line in unit of $f_{\text{lim,1}}$, and we apply three values $f_{\text{lim,1}}=0.25, 0.5, 1.0$ to investigate its impact on the measurement of linear bias. For galaxies with $z>2.0$, we only impose the flux limit $f_{\text{lim,1}}$ on [OIII] emission, since \glc\ does not yet give reliable predictions for [OII] emission. In Figure \ref{fig:radec}, we present the angular distributions of a subsample of our mock galaxies for three different line flux limits (for both H$\alpha$ and [OIII] lines): $5\times10^{-17}\mathrm{erg}/\mathrm{s}/\mathrm{cm}^{2}$ (3D-HST 5$\sigma$ depth), $10^{-16}\mathrm{erg}/\mathrm{s}/\mathrm{cm}^{2}$ (Reference \romanM\ HLSS 6.5$\sigma$ depth), and $2\times10^{-16}\mathrm{erg}/\mathrm{s}/\mathrm{cm}^{2}$ (\euclid\ 3.5$\sigma$ depth). It clearly shows the impact of the survey depth on the measured galaxy distribution that traces the underlying matter distribution. The galaxy number density is very sensitive to the line flux limit. The Reference \romanM\ HLSS (middle panel) is deep enough to trace the underlying matter distribution in detail.

\begin{figure*}
\centering
\includegraphics[width = 7.5in]{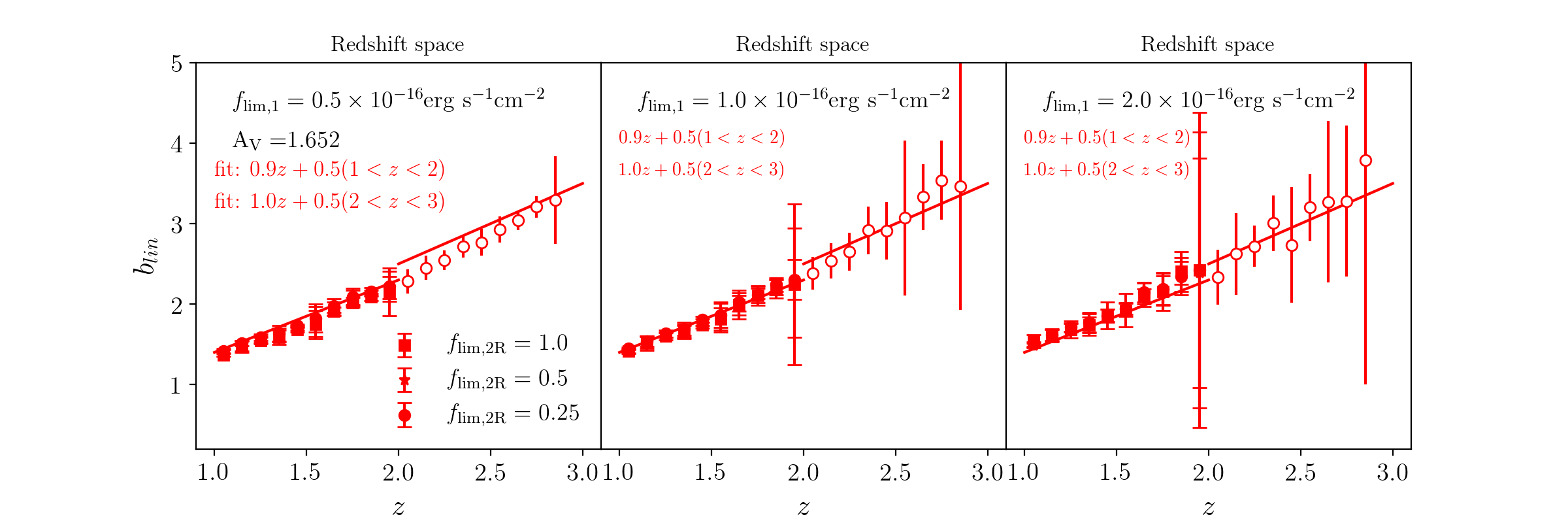}
\label{fig:bias}
\caption{Linear bias of ELGs from \romanM\ HLSS as a function of redshift, under different selections of emission line flux. The filled and empty symbols are results from for H$\alpha$ ($z<2.0$) and [OIII] ($z>2.0$) galaxies respectively. The lines are fitted using galaxies with $f_{\text{lim,1}}=1.0\times10^{-16}\mathrm{erg}/\mathrm{s}/\mathrm{cm}^{2}$ and $f_{\text{lim,2R}}=1.0$, but is also able to describe galaxies with other selections.
The large error bars at $z\sim 2$ and $z\sim 3$ indicate where H$\alpha$ and [OIII] fall outside of the \romanM\ wavelength range of 1-1.93$\mu$m respectively.}
\end{figure*}

We measure the galaxy bias by comparing the galaxy correlation function with the matter correlation function using $\xi_{gg}(r)=b^{2}(r)\xi_{mm}$. The galaxy bias estimated from this method is a function of scale. However, it is close to a constant at scales from 10 to 50 $h^{-1}$Mpc. At smaller scales, the non-linearity of dark matter dynamics becomes significant, while at large scales the bias deviates from a constant due to a combination of factors including RSD effect, sample variance and mode coupling. Therefore the measurement of correlation function at 10 to 50 $h^{-1}$Mpc is best for providing reasonable estimates of the galaxy linear bias. In Figure \ref{fig:bias}, we present the measurement of bias as a function of redshift, for galaxy samples selected with different flux limits. We find that the flux limits and dust models only have mild effects on the linear bias, and the resultant measurement is close to a linear function of redshift, $b_{\text{lin}}=az+b$, with parameters $a$ and $b$ determined through a likelihood analysis. For H$\alpha$ galaxies, the bias measurement can be approximated by $b_{\text{lin}}=0.9z+0.5$, while for [OIII] galaxies, the bias is close to $b_{\text{lin}}=z+0.5$. These linear fits are roughly within $1\sigma$ of measurements using different sample selections, see Fig.\ref{fig:bias}. More details of the measurement are described in \cite{Zhai_2021b}.

\section{Simulation of Grism Data}
\label{sec:grism_sim}

\begin{figure*}
\centering
\includegraphics[width = 4.5in]{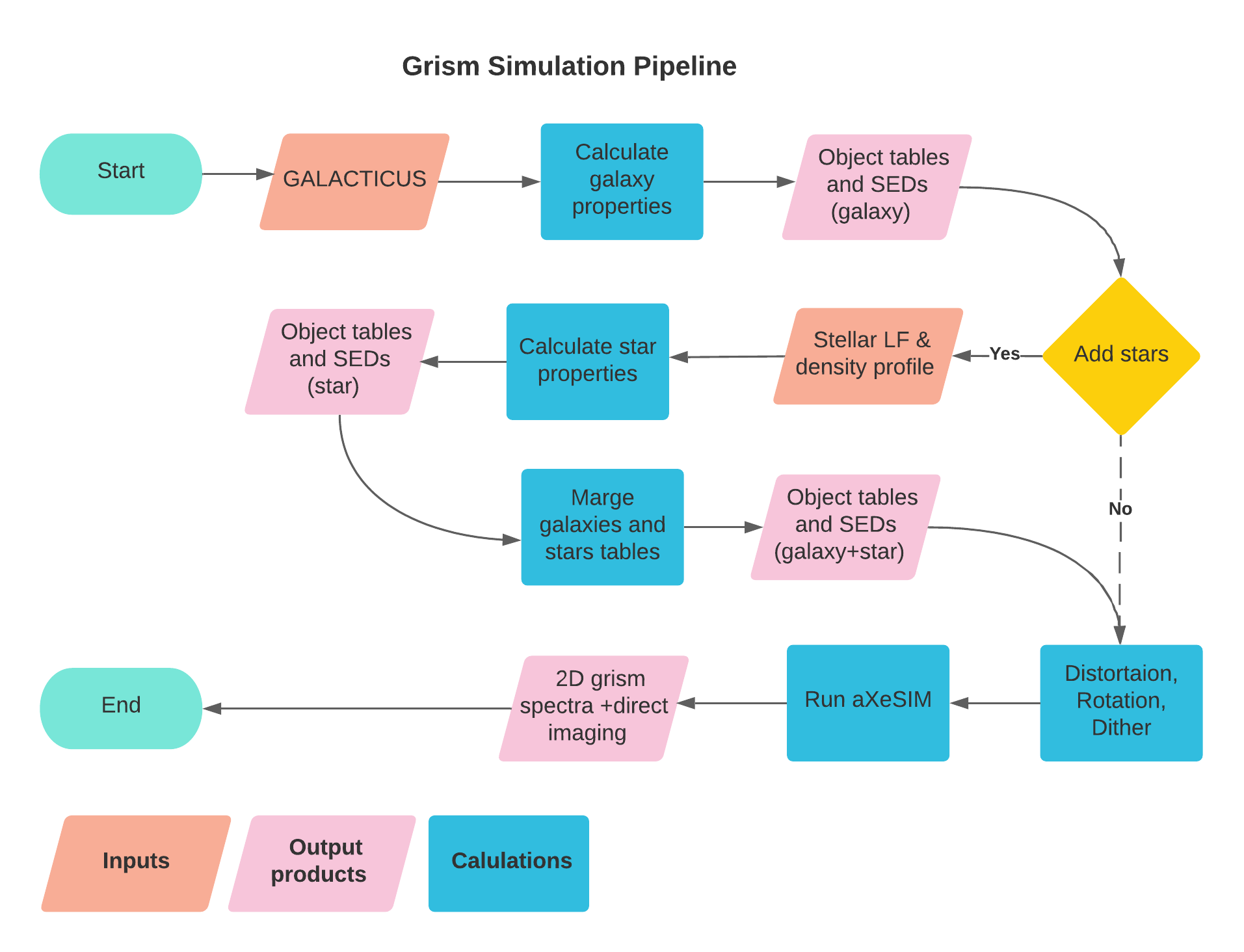}
\label{fig:grism_sim_flowchart}
\caption{Illustration of the steps in our \romanM\ grism observation simulations. In this flow chart, we display the input catalogs and output products with orange and pink  parallelograms, respectively. We also represent various calculation processes in blue rectangles. As described in Section \ref{sec:grism_sim}, in addition to the input galaxy mock catalog produced using Galacticus, the pipeline allows the addition of stars to the  simulated images.  }
\end{figure*}

We have developed a pipeline to simulate grism spectroscopic observations for the \romanM\ HLSS. \romanM\ grism enables spectroscopy at $R=460\lambda/\mu$m for $\lambda=1-1.93\mu$m. Using the pipeline, we have produced simulated data products that are representative of the planned HLSS data, for use by the Science Investigation Team (SIT) and the community. Fig.\ref{fig:grism_sim_flowchart} illustrates an outline of the pipeline and products.

The simulations cover an area of 4 deg$^{2}$ over the redshift range of $0<z<3$. 
By incorporating the survey parameters of the planned HLSS, such as detection limits, exposure times, roll angles, and dithering, the simulation products have been created to resemble as closely as possible the future observations. In addition to the grism slitless spectroscopy, we also simulate the direct imaging that is used to locate the spectra.

\subsection{Motivation behind the grism simulation }
Slitless grism spectroscopic surveys, including HLSS, can suffer from incompleteness and low purity for a variety of reasons. Sources that we fail to detect because their fluxes are close to the detection limit of the survey will result in incompleteness. Furthermore, some sources may be  missed due to their flux and/or emission lines blending with those of nearby objects. In addition to completeness, the fraction of sources with correctly identified emission lines, known as purity, is also a crucial parameter in galaxy redshift surveys. We have designed the mock slitless spectroscopy observations so that they can be used to measure the completeness and purity for the \romanM\ HLSS (see HLSS6 science requirements in section  \ref{subsec:sciencegoals}). 
 
\subsection{Simulation inputs}
To create a realistic simulation of the sky, the pipeline uses input galaxies and stars with a wide variety of physical properties.
We use the aXeSIM software (Kuemmel et al. 2007) to generate a synthetic grism spectrum for each source.  AXeSIM is a simulation software package developed in support of {\it Hubble Space Telescope} slitless spectroscopy modes. It can, however, also simulate spectral images for other slitless spectroscopic instruments, including the \romanM\ Wide Field Instrument (WFI).
AxeSIM requires tables of sources with information about their coordinates, brightness, shapes, redshifts and spectra.  In addition, aXeSIM requires configuration files describing the trace and dispersion solutions for the instrument.

Here we describe the assumptions we made for the creation of the input tables of sources and more details on how we simulate the grism spectra.

\begin{itemize}

\item Galaxies: We use the \romanM\ HLSS 4 deg$^{2}$ galaxy mock catalog with SEDs (see Sec.\ref{sec:mock}) as the input catalog, from which we select galaxies spanning the redshift range of 0 to 3. 
The input galaxy catalog includes galaxy coordinates, sizes, brightness, and redshifts -- we use these to create the galaxy object tables that later become the inputs to the grism simulation machinery. 
\begin{itemize}
\item Coordinates : The sky coordinates  (Ra,Dec) in the input catalog are converted to the detector coordinates (X,Y) on each Sensor Chip Assembly (SCA).

\item Inclination angle: We assume random inclination angle for each galaxy. 

\item Size: The input catalog provides galaxy sizes in terms of disk and spheroidal radius. These sizes, along with the random inclination angle, are used to calculate the galaxy morphology, in terms of semi-major (a) and semi-minor (b) axes. To avoid including very thin galaxies in our sample, we limited our sources to having a major to minor axis ratio $a/b=1-4$ (e.g., see Fig.~11 in \citealt{odewahn_1997}). It is noteworthy that we had to rescale the sizes to improve agreement with observations. Using data from the WFC3 Infrared Spectroscopic Parallel survey (WISP, \citealt{atek_2010,Colbert_2013}), we examined the size distribution of galaxies in the galaxy mock catalog. We compared the normalized distributions of the semi-major sizes of mock galaxy catalog and WISP galaxies. To make the comparison, we only used mock galaxies down to the same magnitude limit as the WISP sub-sample (i.e., H-band $= 25$ AB magnitude). Although the normalized size distribution of the mock catalog has the same peak as the WISP galaxies, the mock catalog contains more large galaxies.  In order to make the distributions match, we need to rescale the size of large mock galaxies, semi-major$ >2.5$ pixel (i.e., $0.275 \arcsec$),  by a factor of $1/2$. We apply the same rescaling factor to the semi-minor sizes in the mock galaxy catalog. 
Incorporating these rescaled sizes, we then use two-dimensional Gaussian profiles as image templates to model the light profiles of galaxies in our final images.  

\item Brightness:  We apply a magnitude cut of $m=28$ AB magnitude in the \romanM\ H-band filter (F158). This value corresponds to the 5$\sigma$ point source detection limit for a 1 hour exposure time in the H-band direct imaging. 

\item SED: In addition to the galaxy properties, we also use the SEDs (see Sec.\ref{sec:mock} for more details) that accompany the input catalog, as the spectral template for each object. We provide a detailed description of how these SEDs and their various components, continuum and emission lines, are constructed in Sec.\ref{sec:mock}. We truncate the SEDs to span an observed wavelength range of $\lambda=10000 -20000$ \AA\ to match the wavelength range of \romanM\ WFI grim observations. 

\end{itemize}
\end{itemize}

\begin{itemize}
\item Stars: For a realistic sky scene, we also incorporate stars in our simulation. We use the model presented in \citet{chang_2010} for the number of stars per unit of apparent magnitude, $m$, within a certain solid angle ($\Delta\Omega$) as follows:

\begin{equation}
\label{equ:num_star} 
\phi(m) = \Delta\Omega \int \psi(M,r)\,n(r)r^{2}{\rm d}r
\end{equation}

where $\psi(M,r)$ is the local stellar luminosity function taken from \citet{Just_2015}. The stellar density profile, $n(r)$, is based on a model presented in \citet{Juric_2008} as the sum over disk and spheroidal halos. It is a good approximation to neglect the angular variation in stellar density over 4 deg$^2$. Note that we only use a single stellar type of G0V for the stellar spectral templates for simplicity. In order to model starlight profiles, we use PSF images as the image templates. Models of the PSFs are constructed using the WebbPSF Python package from STScI \citep{perrin_2014}, which transforms models of the telescope and instrument optical state into PSFs taking into account detector pixel scales, rotations, filter profiles, and point source spectra. In order to optimize both the code run time and the final spatial resolution simultaneously, we made the PSFs with an oversample factor of 2 inside a $364 \times 364$ pixel$^{2}$ box. 

\item Roll angle and dither: In HLSS, spectra will be obtained at different roll angles. This will help resolve possible ambiguity in the correct source to associate with pixels illuminated by overlapping spectra, and thus mitigate the incompleteness caused by spectral confusion and emission lines obscured by nearby neighbors. The simulated products assume 4 different roll angles: 0, 5, 170, and 175 degrees. In addition, to cover the gap between the detectors, we include a two point dither pattern, $\Delta X(\Delta Y)=0 \ \& \ \Delta X(\Delta Y)=1/4$ SCA size (where "SCA" denotes the area covered by one H4RG detector), at each roll angle. 

\item Background and exposure time: For the grism observation background we use a single value of 0.57 e/sec/pix, based on a zodiacal background calculation representative of the HLSS sky coverage. The direct imaging filter has a narrower bandpass, so for the direct image simulations we use a lower value of 0.25 e/sec/pix for the  background. Note that these are typical values and that the sky may vary within the HLSS area. For the simulated grism and direct imaging exposures, we use 301 sec and 141 sec exposure times, respectively, which are the current survey parameters of the Reference HLSS.
\end{itemize}
   
In our current simulations, we have not reproduced the precise pattern of observations represented by the illustrative survey strategy in Fig.~\ref{fig:tile}. However, our simulations of four roll angles and a 1/4 SCA dither can be used to approximate the outcome of this strategy and to examine some alternatives.  Full optimization that uses the grism simulations to test the impact of different strategies on survey completeness, contamination, and uniformity is a goal for future work.

\subsection{Simulation overview}
To run the full simulation, we first divide the 4 deg$^{2}$ area into 15 regions each covering, at most, one \romanM\ field of view. The pipeline simulates one of these regions at a time, with all WFI 18 detectors in their proper configurations. The pipeline first builds a large input object table for each region that includes all the sources contained within that region. Source coordinates in this input table are in the focal plane local coordinate system of the WFI. Next, the pipeline creates 18 smaller lists associated with 18 detectors and converts the source coordinates to the local detector coordinate system ((X,Y) in SCA coordinates, see above). The input list for each individual detector is big enough that it will not miss any sources even when the detector FOV is rotated.

The pipeline then  runs aXeSIM to simulate the images and spectra. We note that aXeSIM has some limitations that we had to solve for these simulations. First, aXeSIM runs one detector at a time. Therefore, we parallelize the pipeline so that all 18 detectors run simultaneously. Second, aXeSIM disperses light only along the x-axis. However, \romanM\ WFI images will be dispersed along the y-axis. To overcome this limitation, we combine a set of 90 degree rotations in both the input files and final products. Third, aXeSIM does not account for instrument distortion from sky coordinates to detector coordinates. In our input object tables, we add distortion to the coordinates. We then manually add distortion coefficients to the header file of the aXeSIM products. 

The final simulation data products include essentially all of the effects that need to be considered in evaluating plans to extract spectra. However, a few limitations were necessary. In the final simulation products, all detector-level artifacts have already been removed. Therefore, the images do not contain cosmic rays, bad pixels, saturation, or dark currents. Additionally, the products are already flat-fielded. We note that these simulations do not include the 0th and 2nd order spectra, the wavelength dependent PSF, or changes to sky background within the HLSS coverage. Additionally, in these simulations, we had to exclude some very narrow ranges of redshifts from our final products. The WFI sensitivity curve that we used for the simulation goes slightly bluer (9000 \AA) than the blue wavelength cut of our spectral templates at 10000 \AA. As aXeSIM extrapolates the SEDs to the blue edge of the sensitivity curve, we would see unrealistic broad spectral features for galaxies with strong emission lines that are redshifted to 10000 \AA. Thus, we had to exclude such redshifts from our samples. These narrow redshift ranges are  $0.517<z<0.531$, $0.993<z<0.997$, $1.012<z<1.016$, $1.053<z<1.057$ and $1.676<z<1.687$.

\begin{figure*}
\centering
\includegraphics[width = 4.5in]{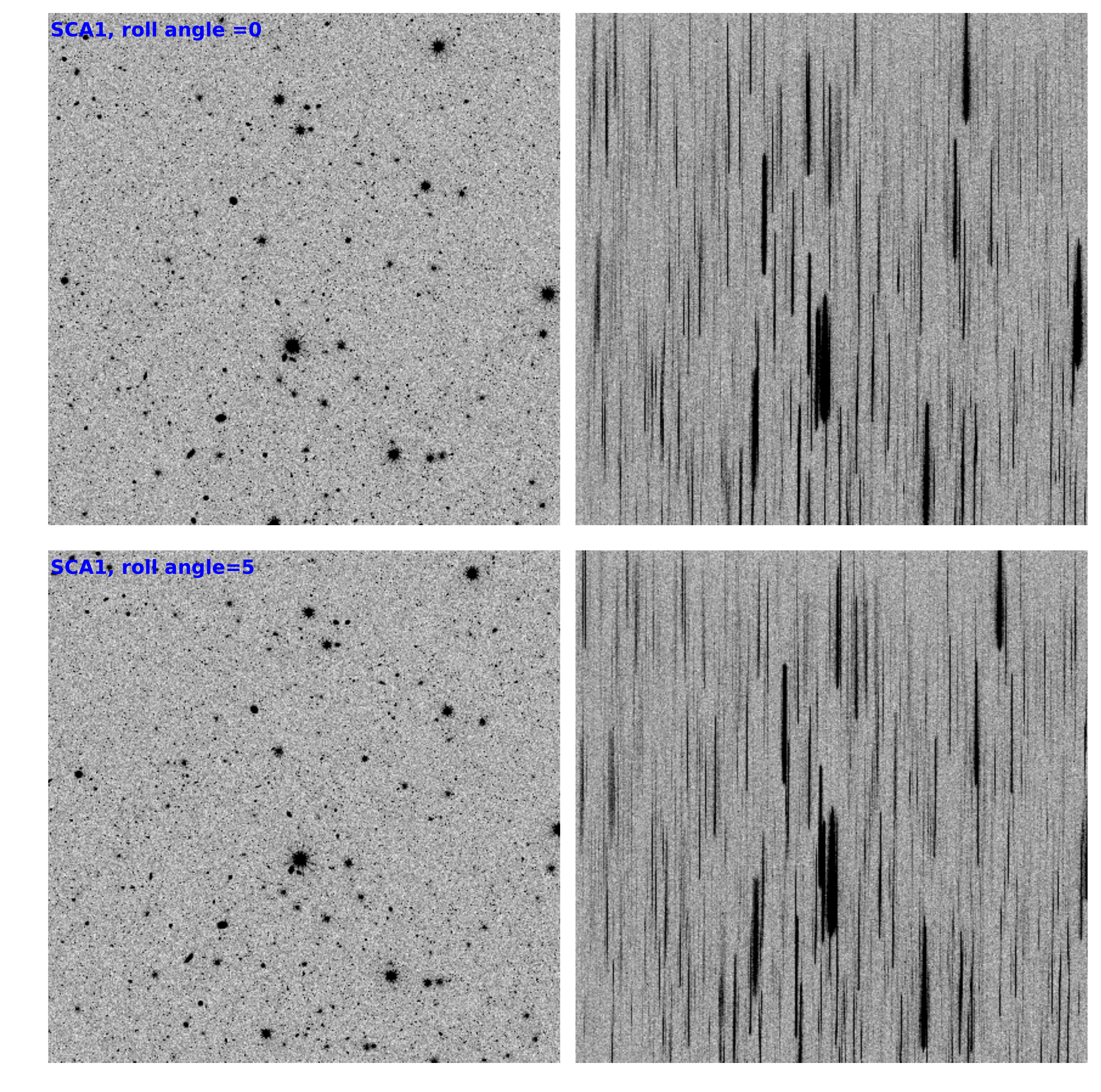}
\label{fig:grssim_example}
\caption{ An example of simulation final products. This example displays a direct image on the left and a 2D slitless dispersed image on the right for detector number 1. The images are shown for two roll angles of 0 and 5 degree.}
\end{figure*}

Our final products will be publicly released on the IPAC website.\footnote{https://roman.ipac.caltech.edu/} The full 4 deg$^{2}$ simulation products are provided in four different roll angles and two different dithering patterns. Each pair of roll angle and dithering includes 18 simulated direct images and 18 simulated 2D slitless dispersed images (i.e., grism spectra). Fig.\ref{fig:grssim_example} displays an example of our simulated direct and dispersed  images for detector number 1 (i.e., SCA1) at two different roll angles of 0 and 5 degree. 

\section{Expected science results}

\subsection{Constraints on cosmic acceleration}
\label{sec:fisher}

In illuminating the unknown nature of cosmic acceleration, we need to measure two free functions of time: the cosmic expansion history $H(z)$, and the growth rate of large scale structure $f_g(z)$. These can tell us whether dark energy varies with time, and whether it is an unknown energy component (e.g., a cosmological constant), or the consequence of the modification of general relativity as the theory of gravity, see e.g., \cite{Wang_2008}.

\romanM\ will have extraordinary capabilities in probing dark energy and constraining gravity.
Here we present the expected fractional errors on $\{H(z)s$, $D_A(z)/s$, $\beta$,  $\overline{f_g(z) G(z)}\}$, where $s$ denotes the BAO scale (the comoving sound horizon at the drag epoch), $D_A(z)$ is the angular diameter distance, $\beta$ is the linear redshift space distortion (RSD) parameter, and $G(z)$ is the growth factor. 
Note that $\overline{f_g(z) G(z)}\equiv f_g(z) G(z) P_0/s^5$ is a derived observable independent of galaxy bias measurements (see \cite{Wang2013}), where $P_0$ is the normalization of the matter power spectrum. We have assumed an observing efficiency of 60\% for all of our forecasts.

We have used the Fisher matrix code calibrated to match analysis results from actual data, see \cite{Wang2013}, and adopted the UNIT input cosmology model as the fiducial model. Our forecast assumptions are conservative: $k_{min}=0.02\,h$/Mpc, $k_{max}=0.25\,h$/Mpc, and 100\% non-linearity (parametrized as $k_*=0.12\, h$/Mpc).
We find that our Fisher matrix code predicts errors that are approximately a factor of 2 smaller than those found by \cite{Zhai_2021} under the same assumptions; this is not surprising since the latter analysis was illustrative using the simplest models for RSD, and not yet optimized via the use of RSD models complex enough to provide high precision.

\cite{Zhai_2021} assumed a line flux limit of $10^{-16}$ergs/s/cm$^2$ at 6.5$\sigma$, which is overly conservative since observers routinely extract and utilize slitless spectra at 5$\sigma$. 
Note that the emission line galaxies have a median radius of $\sim$ 0.25$^{\prime\prime}$ at $z\sim 1.5$, based on WISP data (see Fig.~17 of \cite{Dore18}). 
In this paper, we adopt line flux limits derived for galaxies with radius  0.25$^{\prime\prime}$ at $1.5\mu$m, using the as-built throughputs and wavefront errors of the \romanM\ telescope and instrument. 

The limiting sensitivity to a line emitter depends on reaching a particular SNR. The expected SNR, a property of the galaxy and the instrument, is obtained as the ratio of the signal (calculated using the true line flux) to the noise, with the “signal” defined as the number of photons received from the galaxy. 
At a given signal-to-noise ratio (SNR) cut, the point source sensitivity is significantly better than for galaxies with a finite size, and there is a continuous degradation in sensitivity as the source gets larger. 
A detailed discussion of the SNR for \romanM\ can be found in \cite{Hirata2012}.
The Reference HLSS (see \S\ref{sec:ref-survey}) line flux limit at 1.5$\mu$m, for galaxies with radius 0.25$^{\prime\prime}$, is $\sim 8.5\times 10^{-17}$ergs/s/cm$^2$ at 5$\sigma$, and $\sim 1.0\times 10^{-16}$ergs/s/cm$^2$ at 6.5$\sigma$.

The Reference HLSS is rather minimal; it covers 2000 square degrees to the H$\alpha$ line flux limit of $10^{-16}$ergs/s/cm$^2$ at 6.5$\sigma$ ($\sim 8.5\times 10^{-17}$ergs/s/cm$^2$ at 5$\sigma$), over approximately 6 months of observing time (not including overheads and calibration). 
Table \ref{tab:fisher1} shows the fractional errors on $\{H(z)s$, $D_A(z)/s$, $f_g(z) G(z) P_0/s^5$, $\beta\}$, for the Reference HLSS over 6 months, at the 6.5$\sigma$ and 5$\sigma$ line flux limits respectively.

\begin{table}[]
    \centering
    \begin{tabular}{|c|lc|lc|}
    \hline
  2000 deg$^2$  &$10^{-16}$ergs$/$s$/$cm$^{2}$ (6.5$\sigma$)  & 6 mos 
  & $8.5\times 10^{-17}$ergs$/$s$/$cm$^{2}$ (5$\sigma$)  & 6 mos   \\
   \hline
  $z$  & $H(z)s \hskip 0.25in D_A(z)/s$ \hskip 0.2in $\overline{f_g(z) G(z)}$ & $\beta$ & $H(z)s \hskip 0.25in D_A(z)/s$ \hskip 0.2in $\overline{f_g(z) G(z)}$ & $\beta$
  \\ \hline
    1.0-1.2  & 0.0177  \hskip 0.2in  0.0148 \hskip 0.2in  0.0599 &  0.0329 & 0.0173 \hskip 0.2in 0.0144 \hskip 0.2in 0.0552  &  0.0314
    \\ \hline
    1.2-1.4 &  0.0173 \hskip 0.2in  0.0138 \hskip 0.2in  0.0629 &  0.0364 &  0.0166 \hskip 0.2in 0.0133 \hskip 0.2in  0.0579 &  0.0344
    \\ \hline
    1.4-1.6 &  0.0175 \hskip 0.2in  0.0134 \hskip 0.2in  0.0663 &  0.0407 &  0.0166 \hskip 0.2in 0.0127 \hskip 0.2in  0.0609  & 0.0381
    \\ \hline
    1.6-2.0 &  0.0150 \hskip 0.2in  0.0109 \hskip 0.2in  0.0684 &  0.0416 &  0.0135 \hskip 0.2in 0.0097 \hskip 0.2in 0.0625 & 0.0378
    \\ \hline
    2.0-2.4  & 0.0300 \hskip 0.2in  0.0216 \hskip 0.2in  0.1026 &  0.0955 &   0.0246 \hskip 0.2in  0.0173  \hskip 0.2in 0.0874 & 0.0783
    \\ \hline
    2.4-3.0 &  0.0401 \hskip 0.2in  0.0289 \hskip 0.2in  0.1445 & 0.1466 &   0.0308 \hskip 0.2in 0.0217  \hskip 0.2in 0.1148  & 0.1128
    \\ \hline
    \end{tabular}
\caption{The fractional errors on $H(z)s$, $D_A(z)/s$, $\overline{f_g(z) G(z)}\equiv f_g(z) G(z) P_0/s^5$, and $\beta$, for the Reference \romanM\ HLSS over 6 months (not including overheads and calibration), at the 6.5$\sigma$ and 5$\sigma$ line flux limits respectively. We have assumed an observing efficiency of 60\%.}
\label{tab:fisher1}
\end{table}

\begin{table}[]
    \centering
    \begin{tabular}{|c|l|l|l|}
    \hline
      &  $7\times 10^{-17}$ergs$/$s$/$cm$^{2}$, 4000 deg$^2$, 18 mos &
        $1.5\times10^{-16}$ergs$/$s$/$cm$^{2}$, 10,000 deg$^2$, 10 mos & 
    $2\times 10^{-16}$ergs$/$s$/$cm$^{2}$, 15,000 deg$^2$, 10 mos  \\
 \hline
  $z$  & $H(z)s$ \hskip 0.2in $D_A(z)/s$ \hskip 0.15in  $\overline{f_g(z) G(z)}$ \hskip 0.2in  $\beta$
  & $H(z)s$ \hskip 0.2in $D_A(z)/s$ \hskip 0.15in  $\overline{f_g(z) G(z)}$ \hskip 0.2in  $\beta$ 
  & $H(z)s \hskip 0.2in D_A(z)/s$ \hskip 0.15in  $\overline{f_g(z) G(z)}$ \hskip 0.2in  $\beta$ 
  \\ \hline
    1.0-1.2 &  0.0116 \hskip 0.15in 0.0099  \hskip 0.2in   0.0309 \hskip 0.2in   0.0200&
    0.0084 \hskip 0.15in 0.0071 \hskip 0.2in  0.0271  \hskip 0.2in  0.0158  & 
    0.0077 \hskip 0.15in 0.0064  \hskip 0.2in  0.0259 \hskip 0.2in   0.0148\\ \hline
    1.2-1.4 &  0.0111 \hskip 0.15in 0.0091  \hskip 0.2in    0.0326 \hskip 0.2in  0.0219&
    0.0086 \hskip 0.15in 0.0070 \hskip 0.2in  0.0290  \hskip 0.2in 0.0182  & 
    0.0083  \hskip 0.15in 0.0067  \hskip 0.2in  0.0279 \hskip 0.2in  0.0177\\ \hline
    1.4-1.6 &  0.0110 \hskip 0.15in 0.0085  \hskip 0.2in   0.0346 \hskip 0.2in  0.0242&
    0.0093 \hskip 0.15in 0.0072 \hskip 0.2in 0.0312  \hskip 0.2in 0.0215   & 
    0.0097  \hskip 0.15in 0.0076  \hskip 0.2in 0.0308 \hskip 0.2in  0.0226\\ \hline
    1.6-2.0 &  0.0086 \hskip 0.15in 0.0062  \hskip 0.2in   0.0351 \hskip 0.2in  0.0233&
    0.0096 \hskip 0.15in 0.0073 \hskip 0.2in 0.0338  \hskip 0.2in 0.0254   & 
    0.0113  \hskip 0.15in 0.0089  \hskip 0.2in 0.0355 \hskip 0.2in  0.0299\\ \hline
    2.0-2.4 &  0.0144 \hskip 0.15in 0.0100  \hskip 0.2in   0.0496 \hskip 0.2in  0.0456 &
    0.0219 \hskip 0.15in 0.0165 \hskip 0.2in 0.0673  \hskip 0.2in 0.0701    & 
    0.0229  \hskip 0.15in 0.0176  \hskip 0.2in 0.0697 \hskip 0.2in  0.0733\\ \hline
    2.4-3.0 &  0.0156 \hskip 0.15in 0.0106  \hskip 0.2in   0.0592 \hskip 0.2in  0.0572&
    0.0383 \hskip 0.15in 0.0287 \hskip 0.2in 0.1289  \hskip 0.2in 0.1393  &  
    0.0504  \hskip 0.15in 0.0385 \hskip 0.2in  0.1682 \hskip 0.2in  0.1830\\ \hline
    \end{tabular}
\caption{The fractional errors on $H(z)s$, $D_A(z)/s$, $\overline{f_g(z) G(z)}\equiv f_g(z) G(z) P_0/s^5$, and $\beta$, for an example deep \romanM\ HLSS at $f_{\rm line}>7\times10^{-17}$erg/s/cm$^2$, an example medium \romanM\ HLSS with $f_{\rm line}>1.5\times10^{-16}$erg/s/cm$^2$, and an example shallow \romanM\ HLSS with $f_{\rm line}>2\times10^{-16}$erg/s/cm$^2$. We have assumed an observing efficiency of 60\% and 5$\sigma$ line flux limits.
The estimated observing times do not include overheads and calibration.}
\label{tab:fisher2}
\end{table}

Given the same amount of observing time, the \romanM\ HLSS can cover a wider area at shallower depth, or a smaller area at increased depth. Table \ref{tab:fisher2} shows the fractional errors on $\{H(z)s$, $D_A(z)/s$, $f_g(z) G(z) P_0/s^5$, $\beta\}$, for an example 18 month deep \romanM\ HLSS at $f_{\rm line}>7\times10^{-17}$erg/s/cm$^2$ over 4000 deg$^2$, an example 10 month medium wide \romanM\ HLSS with $f_{\rm line}>1.5\times10^{-16}$erg/s/cm$^2$ over 10,000 deg$^2$, and an example 10 month shallow and very wide \romanM\ HLSS with $f_{\rm line}>2\times10^{-16}$erg/s/cm$^2$ over 15,000 deg$^2$ (the same survey area and line flux limit as \euclid\ but at 5$\sigma$ instead of 3.5$\sigma$). We have assumed an observing efficiency of 60\% and 5$\sigma$ line flux limits. The line flux limits are calculated for a galaxy with half light radius 0.25$^{\prime\prime}$ at 1.5$\mu$m (H$\alpha$ at $z=1.3$). 

Based on our study, the 18 month deep \romanM\ HLSS at $f_{\rm line}>7\times10^{-17}$erg/s/cm$^2$ over 4000 deg$^2$ provides the most compelling stand-alone constraints on dark energy from \romanM\ alone. This provides a useful reference for future optimization of the \romanM\ HLSS given the landscape of results from other projects.

Note that the estimated observing times for the example surveys in Tables \ref{tab:fisher1} and \ref{tab:fisher2} do not include overheads and calibration. These are meant as illustrative examples of what the \romanM\ HLSS could look like.

\subsection{Systematic uncertainties and mitigation strategies}

The systematic uncertainties of the \romanM\ HLSS can be discussed in broad categories: 
(1) Astrophysical and theoretical uncertainties: uncertainties in the modeling of astrophysical and physical processes relevant to BAO/RSD measurements;
(2) Observational uncertainties: systematic effects due to the slitless spectroscopy.  
\subsubsection{Astrophysical and theoretical uncertainties}

The modeling and mitigation of astrophysical and theoretical uncertainties is an extremely active field of study. The main astrophysical uncertainties arise primarily from nonlinear clustering of matter on intermediate and small scales, and the redshift-space distortions (RSD).

The inclusion of intermediate and small scales greatly enhances cosmological constraining power of BAO/RSD, thus the modeling of nonlinear effects is of critical importance. A lot of progress has been made, with the reconstruction of the linear matter density field on intermediate scales being the most effective approach.

RSD is a source of uncertainties on small scales (the random peculiar velocities of galaxies), especially on scales smaller than $\sim$ 5 Mpc/$h$, and the source of signal from cosmic structure growth on large scales (the flattening of 3D galaxy distribution due to the coherent bulk flows). Thus the correct modeling of RSD is essential to meeting the \romanM\ BAO/RSD science goals. This is a research area less advanced than the modeling of nonlinear effects, but progress is being made continuously.

\subsubsection{Observational uncertainties}

\romanM\ spectroscopy suffers from the observational uncertainties due to its slitless spectroscopy and limited spectral coverage.
The slitless nature of the spectroscopy leads to the overlap/blending of spectra from different galaxies.
The limited spectral coverage of 1-1.93$\mu$m leads to the misidentification of emission lines in a given spectrum.
The spectral overlap can be mitigated via the requirement of having at least three roll angles per field. The mitigation of line misidentification is more complex and challenging to address.

As a worked example, \cite{Massara_2020} studied the line confusion in spectroscopic surveys and its possible effect on the BAO analysis. [OIII] emitters are the main target of Roman at z > 1.94. The [OIII] feature on galaxy spectra is a doublet that could be detected by Roman as two separate lines. However, low signal-to-noise spectra might make the detection of the second line very difficult and [OIII] more prone to line confusion. Of particular interest is the possibility of misidentifying H$\beta$ lines as [OIII] features. Indeed, the H$\beta$ line is very close in wavelength to [OIII], so that the two lines cannot be distinguished by photometric measurements. Misidentifying H$\beta$ as [OIII] leads to a wrong inference of the galaxy distance, which will appear 90 Mpc/$h$ closer than its true position. Since the BAO peak present in the galaxy correlation function is at about 100 Mpc/$h$, shifting some galaxies by 90 Mpc/$h$ will lead to an enhancement of the correlation function around the same scale. The interloper-induced bump will convolve with the BAO peak, broadening and shifting it. Misidentifying H$\beta$ as [OIII] introduces a systematic error in the BAO analysis, which might be worrisome but seems to be detectable. Since the shift of galaxies happens only along the line-of-sight, it creates an anisotropic pattern in the galaxy correlation, which should be observed in the quadrupole of the correlation function. We are developing a theoretical model to capture this effect, and allow for its correction.

\subsection{Additional science}

In addition to the standard BAO and RSD measurements from the two-point statistics of the observed galaxy field, the \romanM\ HLSS galaxy samples will enable additional cosmological probes of high redshift cosmology utilizing higher order statistics and small scale clustering. Most of the additional methods either partially or fully rely on the simulation based modelling. The fact that \romanM\ samples are above $z = 1$, where the nonlinearities in the gravitational evolution are still mild, makes the production of the simulations necessary for their science analysis significantly cheaper. The methodology for performing these types of analysis has evolved significantly in recent decade and we expect it to mature even more by the time \romanM\ starts collecting data. Below we will briefly describe the current status quo for a number of key additional probes along with rough projections on how we expect them to perform on \romanM.

\subsubsection{Higher Order Clustering}

An obvious way of going beyond standard BAO/RSD analysis is to analyse higher order correlation functions \citep{1984ApJ...279..499F}. The three-point correlation functions have been measured and analysed since the seventies \citep{1977ApJ...217..385G}. More recently, the three-point functions have been used to measure the BAO signal \citep{2017MNRAS.469.1738S,2018MNRAS.478.4500P}, supplement the two-point function to tighten RSD constraints \citep{2017MNRAS.465.1757G}, constrain neutrino mass \citep{2020JCAP...03..040H}, detect relativistic effects \citep{2019JCAP...04..053D}, and measure primordial non-Gaussianity \citep{2009ApJ...703.1230J,2018MNRAS.478.1341K}.

The most pressing issues for the three-point function analysis are the reduction of the data vector \citep{2018MNRAS.476.4045G,2019MNRAS.484L..29G}, accurate modelling of covariance matrices \citep{2020MNRAS.497.1684S}, and the modeling of the three-point function as a function of cosmological parameters \citep{2018JCAP...04..055L}. On all these issues significant progress has been made in recent years and we expect the methodology to be settled by the time \romanM\ data is delivered for analysis. We expect higher order clustering to significantly enhance the dark energy constraints from the \romanM\ HLSS.

\subsubsection{Void Cosmology}

In the last decade cosmic voids, the under-dense regions in the galaxy distribution, have emerged as a robust probe of cosmology --- complementary to traditional analysis. Given its high galaxy number density coupled with a large volume, \romanM\ holds the power to provide strong constraints on cosmology from voids \citep{2015PhRvD..92h3531P,2019BAAS...51c..40P}. 

The \romanM\ high galaxy number density provides access to the hierarchy of voids and sub-voids, allowing measurements of the void-galaxy cross-correlation function (i. e. the void density profile) down to the smallest scales. The study of the void-galaxy cross-correlation function is a rich field of research and has already provided stringent constraints from current data \citep[e.g.][]{2016PhRvL.117i1302H,2017A&A...607A..54H,2020JCAP...12..023H,2020arXiv200709013A,2020MNRAS.499.4140N}, through its use of voids as standard spheres for the Alcock-Paczy\'nski~ test \citep[]{1979Natur.281..358A,2012ApJ...754..109L} and the redshift-space distortion analysis around void centers. Unlocking small scales with \romanM\ will strongly contribute to tighten such constraints. 
Analogously, accessing small scales permits to constrain the void size function (giving void numbers as a function of size) in the new regime of small void radii, so far unexplored by other surveys.

Voids are particularly sensitive to dark energy \citep[e.g.][]{2009ApJ...696L..10L,2012ApJ...754..109L,2012MNRAS.426..440B,2015PhRvD..92h3531P,2019JCAP...12..040V}, modified gravity \citep[e.g.][]{2015MNRAS.454.3357C,2015MNRAS.451.1036C,2019PhRvD..99f3525S,2021MNRAS.504.5021C} and neutrinos \citep[e.g.][]{2015JCAP...11..018M,2016JCAP...11..015B, 2019PhRvD..99f3525S, 2019MNRAS.488.4413K,2019JCAP...12..055S,2021arXiv210205049B,2021arXiv210702304K}. We have measured $\sim$ 90,000 voids in the \romanM\ HLSS galaxy mock (see Sec.\ref{sec:mock}), using the publicly available Voronoi-watershed based void finder VIDE, and will be studying these to forecast cosmological constraints. In the redshift range $z=1-2$, \romanM\ will be a unique survey for void science to investigate cosmology and physics beyond $\Lambda \mathrm{CDM}$.

\subsubsection{Small Scale Clustering}

Perturbation theory cannot be used to predict clustering patterns of galaxies on small scales. Uncertainties in how the galaxies populate their host dark matter halos \citep{2018PhR...733....1D} complicate theoretical modelling in addition to nonlinear gravitational evolution. A number of recent works have tried to extract cosmological constraints from small scale galaxy clustering despite these difficulties. \citet{2014MNRAS.444..476R} were able to constrain the growth rate of structure with a 2.5 percent precision using CMASS galaxy sample, better than the conventional large scale RSD analysis of the same sample by a factor of two. Subsequent similar works have improved the modelling \citep{2015ApJ...810...35K,Wang2017,2019ApJ...884...29N,2019ApJ...874...95Z}.

Cosmological analysis of the \romanM\ small scale clustering patterns is a very promising avenue, since the high density of its tracers will allow for high precision measurements. The modelling of \romanM\ measurements will likely require a suite of high resolution simulations and a well tested simulation based modeling techniques \citep[similar to e.g.][]{2010ApJ...715..104H,2021arXiv210112261L}.

\section{Conclusion}

\subsection{Summary}

We have presented a comprehensive summary of the extensive work on the \romanM\ HLSS by the \romanM\ HLS Cosmology Science Investigation Team. Our work has included 
\begin{itemize}
    \item{Derivations of science requirements from the \romanM\ dark energy science goal and science objectives.}
    \item{Definition of a Reference HLSS meeting all of the \romanM\ dark energy science requirements.}
    \item{Simulation of realistic galaxy mock catalogs and SEDs for galaxies in the Reference \romanM\ HLSS based on Galacticus, a physical model of galaxy formation.}
    \item{Forecast of galaxy number densities and linear biases using the realistic Reference \romanM\ HLSS galaxy mock, verifying that all \romanM\ dark energy science requirements are met.}
    \item{Forecast of dark energy constraints from the Reference \romanM\  HLSS, as well several possible alternative surveys with different characteristics.}
    \item{Pixel-level grism simulations for the Reference \romanM\ HLSS using our realistic \romanM\ galaxy mock and SEDs as input.}
    \item{Study on the observational systematics for the \romanM\ HLSS.}
    \item{Study on the additional science from the \romanM\ HLSS, utilizing higher-order statistics, void statistics, and small scale clustering, to significantly enhance dark energy constraints from \romanM.}
\end{itemize}

In order to make this paper a useful reference to others, we have included summaries of previous work, in addition to presenting new results obtained as part of this paper.
Note that the Reference HLSS is a worked example of the High Latitude Wide Area Spectroscopic Survey on \romanM. The actual survey that \romanM\ will execute will be defined in an open community process prior to launch, taking into consideration the landscape of dark energy projects and their synergies.

\subsection{Public resources for further work}

This paper provides a comprehensive description of work on the \romanM\ HLSS, with useful results that can help facilitate further work. A companion paper on the \romanM\ HLIS is in preparation \citep{Hirata2021}. 
\cite{Eifler21a} presents multi-probe strategies of cosmology with \romanM.
\cite{Eifler21b} contains a study on the synergies in cosmological studies of \romanM\ with the Rubin Observatory Legacy Survey of Space and Time. A paper considering synergies of Roman with CMB lensing from Simons Observatory is in preparation \citep{Wenzl21}.

We plan to make all of the simulated \romanM\ data available to the public.
The galaxy catalogs and SEDs will be hosted by the NASA IRSA archive at IPAC, jointly presented by the \romanM\ Science Support Center (SSC) at IPAC and IRSA.
The grism simulations will be hosted by the \romanM\ SSC website.

The \romanM\ Exposure Time Calculator \citep{Hirata2012} is available on the \romanM\ SSC website, at \url{https://roman.ipac.caltech.edu/sims/tools/wfDepc/wfDepc.html}.

\acknowledgments
{\bf Acknowledgments:} 
 This work is supported in part by NASA grant 15-WFIRST15-0008, Cosmology with the High Latitude Survey \romanM\ Science Investigation Team. This work was done in part at the Jet Propulsion Laboratory, California Institute of Technology, under a contract with the National Aeronautics and Space Administration.
 We thank Jeff Kruk for cross-checking the \romanM\ ETC and input on the \romanM\ grism simulations, James Rhoads for communications on Reference \romanM\ HLIS+HLSS sky background assumptions, and Lukas Wenzl for helpful comments on a draft of the paper.

\appendix
\section{Modeling galaxies with \glc}

To construct simulated galaxy surveys built on top of cosmological, dark matter-only N-body simulations we make use of the \glc\ semi-analytic model of galaxy formation \citep{benson_galacticus:_2012} to predict galaxy properties. We do not attempt to give a complete description of the physical motivations for the ingredients of the model, which mostly follow standards that have been established in other semi-analytic models, but rather focus on the specifics of the implementation in \glc. The reader is referred to \cite{baugh_primer_2006} and \cite{benson_galaxy_2010} for discussions of the physical motivations for semi-analytic models. \glc\ is designed to be highly modular and flexible. As such, the description given here refers to the specific configuration of \glc\ used to generate the simulated surveys described in this paper.

\glc\ operates on merger trees of dark matter halos (in this case extracted from an N-body simulation as discussed in \S\ref{sec:ngal}). A galaxy is potentially formed within each branch of each merger tree, and is defined by a set of properties which we describe below. Most of these properties are evolved according to a set of differential equations which are detailed in \S\ref{galacticus:differential}. This differential evolution is sometimes interrupted by impulsive events (such as galaxy mergers) as described in \S\ref{galacticus:impulsive}. Finally, some properties (such as galaxy sizes) are determined under assumptions of equilibrium as described in \S\ref{galacticus:equilibrium}.

The \glc\ model contains several free parameters. The values of these parameters have been chosen by manually searching the parameter space and seeking models which provide a reasonable match to a variety of observational data (for example, the $z = 0$ stellar mass function of galaxies; \citealt{li_distribution_2009}) as described in \cite{benson_galacticus:_2012} for example.

\subsection{Properties of Halos and Galaxies}\label{galacticus:components}

In \glc\ each node\footnote{A ``node'' here corresponds with a dark matter halo identified in a dark matter-only cosmological N-body simulation.} of the merger tree is represented by a set of physical components, each of which has one or more properties associated with it. Together, these components, which we will detail below, describe the dark matter halo and the galaxy associated with that node of the tree.

The dark matter halo of each node is modeled with a spherical, NFW profile \citep{navarro_universal_1997}, with virial density contrast given by the spherical collapse model \citep[e.g.][]{percival_cosmological_2005}. The hot atmosphere of gas which fills each dark matter halo is modeled as a spherical, isothermal, $\beta$-profile, with core radius equal to $0.3$ times the virial radius, and with temperature equal to the virial temperature of the halo. The disk of each galaxy is modeled as a razor-thin, exponential disk in which gas and stars share the same radial scale length. A galaxy may also have a spheroid component, modeled as a Hernquist sphere \citep{hernquist_analytical_1990} in which gas and stars share the same radial scale length. Finally, each galaxy may contain a single supermassive Kerr black hole \citep{bardeen_kerr_1970} at its center.

Merger trees are initialized by populating each progenitorless node with a mass of primordial gas equal to the universal baryon fraction multiplied by the dark matter-only simulation halo mass for halos existing prior to reionization at $z=10.5$, or whose virial velocities exceed 35~km/s. For lower velocity halos post-reionization ($z<10.5$) we assume that no gas is accreted. Masses of all other components are set to zero. Scale lengths of the NFW profiles are set using the concentration-mass-redshift relation of \cite{gao_redshift_2008}, while spin parameters \citep{peebles_origin_1969} are assigned using the approach of \cite{cole_hierarchical_2000}. Specifically, a spin is selected at random from the cosmological distribution of \cite{bett_spin_2007}. This spin is assumed to remain unchanged until a halo has doubled in mass, at which time a new spin is selected at random from the same distribution. Halo masses and scale radii are interpolated linearly in time along each branch of the tree, while halo spins are held constant between snapshots\footnote{Note that this means that the halo angular momentum will evolve as the mass and energy of the halo changes in time.}. When a halo merges into a large halo, becoming a subhalo, we assign orbital parameters by drawing at random from the cosmological distribution measured by \cite{benson_orbital_2005}, and assign a time-until-merging for the subhalo using the model of \cite{jiang_fitting_2008} (boosted by a factor 2.25). After this time-until-merging has passed any galaxy in the subhalo is merged with the central galaxy of the host halo as described in \S\ref{galacticus:impulsive} below.

\subsection{Differential evolution}\label{galacticus:differential}

The properties of all components in a node are evolved forward in time according to a set of differential equations. Primordial gas from the intergalactic medium is accreted into each halo at a rate equal to the mass accretion rate of the associated dark mater-only halo scaled by the universal baryon fraction if the halo meets the reionization conditions described above. In this way, any mass growth of a halo which is not accounted for through merging of smaller halos is included as accretion. This accreted mass has a specific angular momentum equal to that of the accreted dark matter (itself determined by the halo spin parameter).

The rate at which gas from the diffuse halo cools and flows into the galaxy is computed using a standard cooling radius model \citep[e.g.][]{white_galaxy_1991} from the evolution of a cooling radius, $r_{\rm cool}$. The cooling radius is defined as the radius at which $t_{\rm cool}(r)=\tau_{\rm v}^{1-\gamma}t_0^\gamma$, where $\tau_{\rm v}$ is the dynamical time of the halo, $t_0$ is the age of the Universe, and $\gamma=0.84$ is a tunable parameter of the model. The cooling time is defined as $t_{\rm cool} = (d/2) {\rm k}_{\rm B} T_{\rm v} n_{\rm t}/\Lambda(T_{\rm v},Z_{\rm h})$, where $d=3$ is the number of degrees of freedom for the cooling gas, $n_{\rm t}$ is the total number density of particles assuming that the gas is. fully ionized, $Z_{\rm h}$ is the metallicity of the hot gas. The cooling function, $\Lambda(T,Z)$ is computed under the assumption of collisional ionization equilibrium and Solar abundances using the \cloudy\ code\footnote{Specifically, we tabulate cooling rates on a grid of temperatures and metallicities. We generally use \cloudy's Solar abundance ratios, with the exception that we also include a zero metallicity case using \cloudy's {\tt primordial} abundances). The helium mass fraction is interpolated linearly with metallicity between \cloudy's primordial and Solar values.} (v13.03; \citealt{ferland_2013_2013}). The angular momentum of this inflowing gas is estimated using the approach of \cite{cole_hierarchical_2000}, and the gas is assumed to settle into a rotationally-supported disk.

We adopt the instantaneous recycling approximation in which we assume that stars with main sequence lifetimes less than around 10~Gyr evolve instantaneously and return some fraction of their mass to the interstellar medium. We adopt a recycled fraction of $R=0.46$ and a yield of $p=0.035$ for a \cite{chabrier_galactic_2001} initial mass function \citep{benson_galaxy_2010}, and further assume that metals are well-mixed into their associated gas at all times. For spheroids we assume that the star formation rate, $\phi_{\rm s}$, exhibits a simple scaling with available gas mass and dynamical time $\phi_{\rm s} = \epsilon_{\rm sf, s} (M_{\rm g, s}/\tau_{\rm s}) \left( 200\hbox{ km s}^{-1}/V_{\rm s} \right)^{\alpha_{\rm sf, s}}$, where $\epsilon_{\rm sf, s}=0.0822$, $\alpha_{\rm sf, s}=1.964$, and $\tau_{\rm s}$ is the dynamical time of the spheroid. For disks, $\phi_{\rm d} = \int_0^\infty {\rm d}R\, 2 \pi R \dot{\Sigma}_{\rm sf, d}(R)$, where $\dot{\Sigma}_{\rm sf, d}(R)$ is the star formation rate surface density and is computed using the expression given by \cite{krumholz_star_2009}. We assume that stellar winds and supernova explosions drive outflows from galaxies at a rate proportional to the rate of star formation with mass-loading factor $\beta = (V_{\rm o} / V)^{\alpha_{\rm o}}$ where $V_{\rm o}=400$ and $85\hbox{ km s}^{-1}$ for disks and spheroids respectively, and $\alpha_{\rm o}=2.1$ and $1.2$ for disks and spheroids respectively. Gas outflowing from the galaxy is assumed to be gradually reincorporated into the hot halo \citep[see][for example]{henriques_simulations_2013} at a rate $\dot{M}_{\rm inc} = \alpha_{\rm inc} M_{\rm o} / \tau_{\rm v}$, where $\alpha_{\rm inc}=5.0$ controls the rate of return of outflowed gas, and $\tau_{\rm v}$ is the dynamical time of the halo.

As a consequence of star formation, stellar luminosity is produced at a rate $\dot{L}_{\rm sf, i} = \dot{\phi}_i \Upsilon^{-1}(Z_{{\rm g}, i},t_{\rm obs}-t)$, where $\Upsilon(Z,t_{\rm a})$ is the mass-to-light ratio of a simple stellar population of metallicity $Z$ and age $t_{\rm a}$, and $t_{\rm obs}$ is the age of the Universe at which the luminosity will be observed. Note that many different luminosities---corresponding to different $t_{\rm obs}$ and different filters (and so to different functions $\Upsilon(Z,t_{\rm a})$)---are computed simultaneously. The mass-to-light ratio is computed by integrating the spectral energy distributions of simple stellar populations under the relevant filter transmission curve. These spectral energy distributions are computed using the FSPS code (v2.5; \citealt{conroy_propagation_2009,conroy_fsps:_2010}) to tabulate for a grid of times and metallicities assuming a \cite{chabrier_galactic_2001} initial mass function. When computing $\Upsilon(Z,t_{\rm a})$ we then interpolate in this table as necessary.

We use the stability criterion originally proposed by \cite{efstathiou_stability_1982} to judge whether galactic disks are unstable to the formation of a spheroidal component \citep{ostriker_numerical_1973}. If the disk of a galaxy is deemed to be gravitationally unstable, we allow mass, metals, and angular momentum to be transferred from the disk component to the spheroid component on a timescale comparable to the dynamical timescale of the disk.

For satellite halos, the hot halo can be stripped due to ram pressure from the hot halo of the host. The ram pressure radius is found following \cite{font_colours_2008}, with mass outside this radius being removed on a timescale based on the ram pressure acceleration \citep{roediger_ram_2007}.

The supermassive black hole is allowed to accrete from both the hot gas halo and the interstellar medium of the spheroid component at rates proportional to the Bondi rate, but limited to the Eddington rate. Bondi accretion rates are boosted by factors of $6.3$ and $4.7$ for accretion from the hot halo and the ISM respectively. For accretion from the hot halo we assume accretion can only occur from the fraction of the hot halo mass actually in the hot mode. We do not explicitly model whether halos are undergoing hot or cold mode accretion, and so instead impose a simple transition from cold-mode to hot-mode behaviour at the point where a halo (were it in the hot mode) is able to cool out to the virial radius \citep{benson_cold_2010}.

As discussed in detail by \cite{begelman_accreting_2014}, accretion flows with accretion rates close to the Eddington limit will be radiatively inefficient as they struggle to radiate the energy they release, while flows with accretion rates that are much smaller than Eddington ($\dot{M}_{\rm acc} < \alpha^2 \dot{M}_{\rm Edd}$, where $\alpha\sim 0.1$ is the usual parameter controlling the rate of angular momentum transport in a \cite{shakura_black_1973} accretion disk) are also radiatively inefficient as radiative processes are too inefficient at the associated low densities to radiate energy at the rate it is being liberated. Therefore, accretion disk structure is assumed to be a radiatively-efficient, geometrically thin, \cite{shakura_black_1973} accretion disk if the accretion rate is between $0.01\dot{M}_{\rm Edd}$ and $0.3\dot{M}_{\rm Edd}$, and an advection dominated accretion flow (ADAF) otherwise \citep{begelman_accreting_2014}. For thin disks and high-accretion rate ADAFs the radiative efficiency is given by $\epsilon_{\rm rad}=1-E_{\rm ISCO}$, where $E_{\rm ISCO}$ is the specific energy of the innermost stable circular orbit (in dimensionless units) for the given black hole spin. For low accretion rate ADAFs the radiative efficiency is matched to that of the thin disk solution at the transition point ($0.01\dot{M}_{\rm Edd}$) and is decreased linearly with accretion rate below that. For the jet efficiency in thin accretion disks we use the results of \cite{meier_association_2001}, interpolating between their solutions for Schwarzchild black holes and rapidly rotating Kerr black holes. For the case of ADAF accretion flows we use the jet efficiency computed by \cite{benson_maximum_2009}.

We model quasar-mode feedback by allowing the black hole to drive a wind from the spheroid component following the model of \cite{ostriker_momentum_2010}, and the assumption that energy deposition into the spheroid results in a wind carrying a mass flux of $\dot{M}_{\bullet, \rm qsr} = \epsilon_{\rm s} L_{\rm wind}/V_{\rm s}^2$, where $\epsilon_{\rm s}=0.01$, when the density of the ISM in the spheroid exceeds the critical value defined by \cite{ostriker_momentum_2010}. We model radio-mode feedback as jet power deposition into the hot halo. We assume that energy is deposited by the jet into the hot halo at a rate $P_{\rm rdo} = \epsilon_{\rm jet} \dot{M}_\bullet {\rm c}^2$, and that this causes a reduction in the mass inflow rate of $P_{\rm rdo}/V_{\rm v}^2$. Inflow rates of metals and angular momentum are reduced proportionally. If the inflow rate is reduced to zero, any remaining jet power is used to push material out of the hot halo at a rate $\dot{M}_{\bullet, \rm rdo} = P_{\rm rdo}/V_{\rm v}^2 - \dot{M}_{\rm inf,0}$, where $\dot{M}_{\rm inf,0}$ is the rate of inflow in the absence of any radio-mode feedback. 

The resulting network of differential equations described above are solved numerically using a backward differentiation formula integrator, with adaptive timestepping chosen to ensure a fractional precision of 1\% in all evolved quantities\footnote{We set small, absolute precisions for all evolved quantities to prevent the differential evolution requiring arbitrarily small timesteps in regimes where some parameter values become extremely small.}. We evolve all properties of all components in a node simultaneously, but evolve only one node at a time. The order in which nodes are evolved is determine by the causal connections between them (i.e. progenitor halos must be evolved before their descendants, while satellites must be evolved in lock-step with their centrals since their properties are mutually interdependent).

The differential equation networks include some cross-terms between nodes (e.g. due to ram pressure stripping). As we evolve only one node at a time these cases must be treated separately. Our approach is to accumulate any mass, metals, or angular momentum transferred from a halo being evolved to another halo over some timestep, $\Delta t$. After that timestep, differential evolution is paused, the accumulated quantities are transferred to the receiving node, and differential evolution resumes (possibly in another node, possibly the node which just received the accumulated properties, according to the rules described above). The size of the timestep, $\Delta t$, is chosen to be sufficiently small to give converged results.

\subsection{Impulsive Evolution}\label{galacticus:impulsive}

In addition to the differential evolution described above, we interrupt this evolution when two galaxies merge. Galaxy properties are changed instantaneously during this event, after which differential evolution resumes. The effects of galaxy mergers are described below. Mergers between galaxies are classified as major if $M_{\rm sat} > f_{\rm major} M_{\rm cen}$ where $M_{\rm sat}$ and $M_{\rm cen}$ refer to the total galactic baryonic mass of the satellite and central galaxies respectively, and $f_{\rm major}=0.57$. In a major merger, all mass in the central galaxy galactic disk is transferred to the spheroid while in a minor merger the central galaxy galactic disk remains in place. In either case, the mass of the satellite galaxy is transferred to the central galaxy spheroid after which the satellite galaxy no longer exists as a distinct entity.

Any remaining hot halo and outflowed gas associated with the satellite galaxy halo is transferred to the central galaxy halo, with the satellite gas assumed to have a specific angular momentum relative to that of the central galaxy halo (which is proportional to the virial specific angular momentum of that halo). The supermassive black holes in the two merging galaxies are assumed to also merge instantaneously. The mass of the final black hole is taken to be the sum of the masses of the two initial black holes (i.e. we ignore any loss of mass into gravitational radiation, which are typically a few percent; \citealt{rezzolla_final_2008}). The spin of the final black hole is computed using the fitting function given by \cite{rezzolla_final_2008} assuming that the initial spins of the two black holes and their orbital angular momentum are all parallel (i.e. $\cos \alpha = \cos \beta = \cos \gamma = 1$ in the notation of \citealt{rezzolla_final_2008}).

\subsection{Equilibrium Determined Properties}\label{galacticus:equilibrium}

Galaxy sizes are determined on the assumption that both disk and spheroid are in equilibrium in the gravitational potential given their angular momenta, following the general approach of \cite{cole_hierarchical_2000}. The radial sizes of the disk and spheroid components are determined from their angular momenta (actually the pseudo-angular momentum in the case of the spheroid, defined below) and the assumption of rotational support in the gravitational potential created by the combined baryonic and dark matter mass distributions in the halo, with the dark matter profile being allowed to respond to the (adiabatic) growth of the galaxy. Sizes are assumed to adjust instantaneously to any changes in angular momentum or gravitational potential, and so are always in equilibrium. While model spheroids are nominally pressure supported we solve for their sizes by defining a pseudo-angular momentum, which is the angular momentum that the spheroid would have if it were rotation supported at the same size. The model of \cite{cole_hierarchical_2000} accounts for adiabatic contraction of the dark matter halo. We also include this effect, but adopt the updated model of \citep{gnedin_response_2004}. During galaxy mergers, the pseudo-angular momentum of any newly-formed spheroid is computed using the method proposed by \cite{cole_hierarchical_2000}.

\end{document}